\documentclass[12pt]{ar2e}
\usepackage{epsfig}
\begin{document}

\input psfig.sty

\def\araa{{\it Ann. Rev. Astron. Astrophys.}}
\def\apj{{\it ApJ}} 
\def\apjl{{\it Ap. J. Lett.}}
\def\apjs{{\it Ap. J. Suppl.}}
\def\mnras{{\it MNRAS}}
\def\aa{{\it Astron. Astrophys.}}
\def\aap{{\it Astron. Astrophys.}}
\def\aj{{\it Astron. J.}}
\def\pasp{{\it Proceedings of the Astronomical Society of the Pacific  }}
\def\nat{{\it Nature}}
\def\bain{{\it Bull. Astron. Inst. Netherlands}}
\def\apss{{\it Astrophysics and Space Science}}

\def\NL{$2{\times}10^{20}${\rm cm}$^{-2}$}
\def\dndz{$d{\cal N}/dz$}
\def\dnzdz{$d{\cal N}(z)/dz$}
\def\msolar{$M_{\odot}$}
\def\dndx{$d{\cal N}/dX$}
\def\pkts{P_{KS}}
\def\etal{et al.}
\def\H2{$\rm H_{2}$}
\def\dla{damped {\lya}}
\def\hnu{$h{\nu}$}
\def\dla{damped {\lya}}
\def\DLA{damped \lya\ system}
\def\DLAs{damped \lya\ systems}
\def\DLAS{Damped \lya\ systems}
\def\delv{$\Delta$v}
\def\fNX{$f(N,X)$}
\def\Dlam{$\Delta \lambda$}
\def\delvH I{${\Delta v}_{\rm H I}$}
\def\delvMgII{${\Delta v}_{\rm Mg II}$}
\def\Lya{Ly$\alpha$}
\def\lya{Ly$\alpha$}
\def\lamlya{${\lambda_{Ly{\alpha}}}$}
\def\Ts{T$_{s}$}
\def\jnu{$J_{\nu}$}
\def\junit{10$^{-19}$ ergs cm$^{-2}$ s$^{-1}$ Hz$^{-1}$ sr$^{-1}$}
\def\sigv{$\sigma_{v}$}
\def\ndmp{17}
\def\nrun{8,500}
\def\kms{km~s$^{-1}$}
\def\cm2{cm$^{-2}$}
\def\N#1{{N({\rm #1})}}
\def\f#1{{f_{\rm #1}}}
\def\rAA{{\rm \, \AA}}
\def\sci#1{{\rm \; \times \; 10^{#1}}}
\def\ltk{\left [ \,}
\def\ltp{\left ( \,}
\def\ltb{\left \{ \,}
\def\rtk{\, \right  ] }
\def\rtp{\, \right  ) }
\def\rtb{\, \right \} }
\def\ohf{{1 \over 2}}
\def\nohf{{-1 \over 2}}
\def\rhf{{3 \over 2}}
\def\smm{\sum\limits}
\def\perd{\;\;\; .}
\def\cmma{\;\;\; ,}
\def\sgint{\sigma_{int}}
\def\Nperp{N_{\perp} (0)}
\def\intl{\int\limits}
\def\und#1{{\rm \underline{#1}}}
\def\nh1{$N$(H I)}
\newcommand{\tenO}{\mbox{$10^{20} \ {\rm cm^{-2}}$}}
\newcommand{\tenOO}{\mbox{$10^{21} \ {\rm cm^{-2}}$}}
\def\NL{$2{\times}10^{20}$cm$^{-2}$}
\def\NH{$N$(H I)}
\def\pkts{P_{KS}}
\def\eslash{$\rm \acute e$}
\def\Lya{Ly$\alpha$} 
\def\alphenh{[$\alpha$/H]}
\def\MovH{[M/H]}
\def\lya{Ly$\alpha$} 
\def\smpy{M$_{\odot}$ yr$^{-1}$}
\def\smpykpc{M$_{\odot}{\rm \ yr^{-1} \ kpc^{-2}}$}
\def\smpympc{M$_{\odot}{\rm \ yr^{-1} \ Mpc^{-3}}$}
\def\lcdm{$\Lambda$CDM} 
\def\rvec{${\bf r}$}
\def\Rratio{${{\cal R}(v)}$}
\def\ndmp{17}
\def\nrun{8,500}
\def\omgm{$\Omega_{{\rm M}}$}
\def\omgv{$\Omega_{\Lambda} $}
\def\omgas{$\Omega_{g}(z) $}
\def\omstar{$\Omega_{*}(z=0) $}
\def\dv{$\Delta v$}
\def\dvm{$\delta v$}
\def\vc{$V_{rot}(r)$}
\def\vrot{$V_{rot}(r)$}
\def\micron{$\mu$m}
\def\gamd{${\Gamma_{d}(\bf r)}$}
\def\gamdnr{${\Gamma_{d}}$}
\def\peq{$P_{eq}$}
\def\pmin{$P_{min}$}
\def\pmax{$P_{max}$}
\def\taunu{$\tau_{\nu}$}
\def\pgeom{$(P_{min}P_{max})^{1/2}$}
\def\kap{${\kappa ({\bf r})}$}
\def\dNdX{$d{\cal N}/dX$}
\def\dNdXz{$d{\cal N}(z)/dX$}
\def\kapnr{${\kappa}$}
\def\N#1{{N({\rm #1})}}
\def\f#1{{f_{\rm #1}}}
\def\rAA{{\rm \, \AA}}
\def\sci#1{{\rm \; \times \; 10^{#1}}}
\def\ltk{\left [ \,}
\def\ltp{\left ( \,}
\def\ltb{\left \{ \,}
\def\rtk{\, \right  ] }
\def\rtp{\, \right  ) }
\def\rtb{\, \right \} }
\def\ohf{{1 \over 2}}
\def\nohf{{-1 \over 2}}
\def\rhf{{3 \over 2}}
\def\smm{\sum\limits}
\def\perd{\;\;\; .}
\def\cmma{\;\;\; ,}
\def\semic{\;\;\; ;}
\def\sgint{\sigma_{int}}
\def\Nperp{$N_{0}$}
\def\frat{$f_{ratio}$}
\def\intl{\int\limits}
\def\rhodot{$\dot{\rho_{*}}$}
\def\mdot{$\dot{{ M}_{*}}$}
\def\zab{$z_{abs}$}
\def\rhodotz{$\dot{\rho_{*}}$$(z)$}
\def\rhodotacc{$\dot{\rho_{a}}$$(z)$}
\def\rhosz{${\rho_{*}(z)}$}
\def\rhog{${\rho_{g}(z)}$}
\def\rhos{${\rho_{*}}$}
\def\und#1{{\rm \underline{#1}}}
\def\ps{$\dot{\psi_{*}}$}
\def\psav{$<\dot{\psi_{*}}>$}
\def\ms{$\dot{M_{*}}$}
\def\psav{$<$$\dot{\psi_{*}}$$>$}
\def\psavz{${<{{\dot{\psi_{*}}}}(z)>}$}
\def\ciis{C II$^{*}$}
\def\nh{$N$(H I)}
\def\lclos{$\ell_{c}$}
\def\lcav{$<l_{c}>$}
\def\lcr{$l_{cr}({\rm {\bf r}})$}
\def\lcrnr{$l_{cr}$}
\def\jnu{$J_{\nu}$}
\def\knu{$k_{\nu}$}
\def\junit{ergs cm$^{-2}$ s$^{-1}$ Hz$^{-1}$ sr$^{-1}$}
\def\funit{ergs cm$^{-2}$ s$^{-1}$}
\def\fnuunit{ergs cm$^{-2}$ s$^{-1}$ Hz$^{-1}$}

\jname{Annual Reviews of Astronomy \& Astrophysics}
\jyear{2005}
\jvol{}
\ARinfo{1056-8700/97/0610-00}

\title{DAMPED \lya\ SYSTEMS}

\markboth{{Wolfe,} {Gawiser \&} {Prochaska}}{Damped \lya Systems}

\author{ Arthur M. Wolfe
\affiliation{Department of Physics, and 
Center for Astrophysics and Space Sciences, 
University of California, San
Diego, 
Gilman Dr., La Jolla; CA 92093-0424, awolfe@ucsd.edu}
Eric Gawiser 
\affiliation{
NSF Astronomy \& Astrophysics Postdoctoral Fellow, 
Department of Astronomy, Yale University, 
P.O. Box 208101, New Haven, CT 06520-8108, gawiser@astro.yale.edu}
Jason X. Prochaska 
\affiliation{
University of California Observatories/ Lick Observatory,
University of California,
333 Interdisciplinary Science Building, 
Santa Cruz CA 95064, xavier@ucolick.org}
}








\begin{abstract}
{Observations of {\DLAs} offer a unique window on the neutral-gas
reservoirs that gave rise to galaxies at high redshifts. 
This review focuses on 
critical properties such
as the H I and metal content of the gas
and on independent evidence for star formation. 
Together,
these provide an emerging
picture of gravitationally bound objects in which
accretion of gas from the IGM replenishes gas consumed by star formation.
Other
properties such as dust content,
molecular content, 
ionized-gas content, gas kinematics,
and galaxy identifications are also reviewed.
These properties point to a multi-phase ISM in which 
radiative and hydrodynamic feedback
processes
are present. 
Numerical
simulations and other types of 
models used to describe {\DLAs}
within
the context of galaxy formation are also discussed.}
\end{abstract}

\maketitle


\section{WHAT ARE THE DAMPED {\lya} SYSTEMS and HOW ARE THEY FOUND?}

Damped {\lya} systems are a class of QSO absorbers selected for the presence
of H I column densities, \nh $\ge$ 2$\times${\tenO}. This 
criterion differs from those used to find other classes of QSO
absorbers selected on the basis of H I content. The {\lya}
forest absorbers, reviewed in this journal by 
\citet{rauch98},
are
selected for  {\nh} $<$ 10$^{17}$ {\cm2}, while
the Lyman limit systems have 10$^{17}$  $<$ {\nh} $<$
2$\times${\tenO} (Peroux {\etal} 2003b).  
The {\lya}
forest absorbers are optically thin at the Lyman limit, since 
the column density
{\nh} = 10$^{17}$ {\cm2} corresponds to about unity optical 
depth at the Lyman limit. 
Are these absorbers physically different from 
the damped systems or have the column-density
criteria resulted in arbitrary distinctions? In fact there
is a fundamental
difference: 
hydrogen is mainly neutral in damped {\lya}
systems, while it is ionized in all other classes
of QSO absorption systems. This includes
absorbers selected
for the presence of C IV {$\lambda$}{$\lambda$} 1548.1, 1550.7 
resonance-line doublets 
\citep*{sargentsb88},
Mg II {$\lambda$}{$\lambda$} 2796.3, 2803.5
resonance-line doublets 
\citep{steidels92}
and Lyman limit absorption \citep{prochaska99}, which do
not also qualify as {\DLAs}.  

The neutrality of the gas is crucial:
while
stars are unlikely to form out of warm ionized gas, 
they are likely to descend from cold neutral clouds,
which are the precursors of
molecular clouds, the birthplace of stars
\citep{wolfireetal03}.
This property
takes on added significance when it is realized that
the damped {\lya} systems dominate the neutral-gas content
of the Universe in the redshift interval $z$=[0,5],
and at z$\sim$3.0$-$4.5 contain
sufficient mass in neutral gas to account for a significant
fraction of the visible stellar mass in modern galaxies
\citep[e.g.][]{storrielombardiw00}.
This has led to the widely
accepted idea that damped {\lya} systems 
serve as important neutral gas reservoirs for star formation
at high redshifts 
\citep*[e.g.][]{nagaminesh04a}.  
Moreover, as  repositories of significant amounts of metals
the {\DLAs} have been used to trace the age-metallicity
relationship and other aspects of galactic 
chemical evolution 
(\citealt{pettinietal94}; \citealt*{peifh99}; \citealt{pettini04}; 
\citealt{prochaskaetal03b}).  

The purpose of this review is to present an overview of the {\DLAs}.
Current research on the high-redshift Universe is dominated by
surveys that rely on the detection of radiation 
emitted by stars 
\citep[e.g.][]{steideletal03,giavaliscoetal04,dickinsonetal03}
 or ionized gas 
\citep[e.g.][]{rhoadsm01,ouchietal03}.  
By contrast, 
{\DLAs} provide a window on the interplay between neutral gas
and newly formed stars, i.e., the {\DLAs} are the best, perhaps
the only, examples we have of an interstellar medium in
the high-redshift Universe. Consequently, the focus of this
review will be on the 
manner in 
which {\DLAs} 
trace, and play an active role in, 
cosmic star formation
and 
hence 
galaxy formation.  

Throughout this article we adopt a cosmology consistent with
the  {\em WMAP} \citep{bennettetal03} results, ($\Omega_{m}$, $\Omega_{\Lambda}$,$h$)
=(0.3,0.7,0.7).

\subsection{History of {\dla} Surveys}

To understand the significance of damped {\lya} systems
for research in galaxy formation we give a brief
historical perspective.

 The motivation for the first
damped {\lya} survey was to find the neutral-gas disks of
galaxies at high redshifts 
\citep{wolfeetal86}.
Unlike today, 
the cold dark matter (hereafter CDM)
 paradigm of hierarchical structure formation 
(i.e., merging protogalactic clumps)
did not dominate theories of galaxy formation in the
early 1980s. Rather,  
the idea of mature galaxy disks at high redshift fitted in with
the coherent collapse model of 
\citet*[][also \citealt{falle80}]{eggenls62},
which
was highly influential at the time.
Some QSO absorbers with
properties resembling galaxy disks had been found 
at $z$ $<$ 1 through the 
detection of 21 cm absorption either in radio-frequency
scanning surveys 
\citep{brownr73}
or
at the redshifts of
Mg II selected absorbers 
\citep{robertsetal76}.
However, application of these techniques resulted in only a few
detections. While it was unclear whether the 21 cm absorbers
belonged to a new population of objects or were 
rarely occurring oddities, 
the
radio scanning techniques were valuable for successfully detecting
cold, quiescent gas at large redshifts
for the first time. Specifically, 
\citet{brownr73}
and 
\citet{brownm83}
 used this technique to detect
two 21 cm 
lines with FWHM velocity widths, {\delvH I} {$\approx$}
10 and 20 {\kms}. 
The temperature of gas detected in 21 cm
absorption is likely to be
low, because the 21 cm
optical depth $\tau_{21}$ $\propto$ {\nh}/($T_{s}${\delvH I}),
where the hyperfine spin temperature, $T_{s}$,
generally equals the kinetic temperature of the
cold, dense gas detected in 21 cm absorption.


However, the most efficient method for locating quiescent layers of 
neutral gas is through the detection of damped {\lya} absorption
lines. 
In the rest frame of the atom
the absorption
profile of any atomic transition is naturally broadened owing
to the finite lifetime of the upper energy state.
In the rest frame defined by the average velocity
of the gas, the natural 
profile is Doppler broadened  by the 
random motions of the atoms:
the convolution of both effects results in the Voigt
profile 
\citep[e.g.][]{mihalas78}.
Because 
the Doppler profile falls off from the
central frequency, ${\nu}_{0}$, as
exp[$-({\Delta \nu}/{\Delta \nu}_{D})^{2}$]
(where ${\Delta \nu}$ = $|{\nu}$$-$${\nu}_{0}|$ and 
${\Delta \nu}_{D}$=${\sqrt 2}{\sigma_{v}}{{\nu}_{0}}/c$
for an assumed Gaussian
velocity distribution with dispersion $\sigma_{v}$) 
and the natural or ``damped'' absorption profile
falls off from $\nu_0$ like 1/$({\Delta \nu})^{2}$,
at sufficiently
large ${\Delta \nu}$ 
the probability for damped absorption
exceeds the probability
for absorption in 
the Doppler profile.
The frequency intervals in which natural broadening
dominates Doppler broadening are called the
damping  wings of the profile
function. Most atomic
transitions of abundant ions are optically thin in their damping
wings but 
optically thick near the core of the Doppler profile.
The latter transitions  have unit optical depth  
at ${\Delta \nu}_{\tau =1}$ $\propto$
${{\Delta \nu}_{D}}$${\times}[\ln N({\rm X}^{j})]^{1/2}$,
where $N$(X$^{j}$) is the column density of ionic species X$^{j}$.
Such lines are saturated. The reason is that the rest-frame equivalent
width of an absorption line is given by  
$W_{r}$$\equiv$($\lambda$/$\nu$)$\int(1-{\rm exp}(-{\tau_{\nu}})d{\nu}$), 
and therefore
$W_{r}$
is proportional to ${\Delta \nu}_{\tau =1}$. In the case of
lines with unit optical depth near the Doppler
core the line is saturated because
$W_{r}$ is insensitive to the value of $N$(X$^{j}$).
Due to the higher values of {\nh}, 
{\lya} has unit optical depth
in the damping wings at
${\Delta \nu}_{\tau =1}$ $\propto$ [$A_{21}f_{21}${\nh}]$^{1/2}$  
when {\nh} $>$ 10$^{19}$
{\cm2} and $\sigma_{v}$ $<$ 70 {\kms}: 
$A_{21}$ and $f_{21}$ are the
Einstein spontaneous emission coefficient and oscillator
strength
for the {\lya} transition.
In this case, unit optical depth occurs in the
damping wings, and therefore
the equivalent width of a damped {\lya} line is independent
of the velocity structure of the gas
for velocity dispersions within the range detected in 
most QSO absorption systems. As a result, the equivalent
width will be large even when the velocity
dispersion is small.

By the
early 1980s only four {\DLAs} 
had been found. In every case 
they were high column-density systems,
{\nh} $>$ {\tenOO}, which were
found by chance 
(\citealt{beaveretal72,carswelletal75}; 
\citealt*{smithmj79}; \citealt{wrightetal79}).
Although the sample was sparse,
the utility of the
damped {\lya} criterion was demonstrated when 21 cm absorption
at $z$$\sim$2
was detected in 2 of the 3 background QSOs that were radio
sources 
\citep{wolfed79,wolfeb81}.
The narrow line widths, {\delvH I}$\approx$20 {\kms}
(where {\delvH I}=${\sqrt {8{\rm ln}2}}$$\sigma_{v}$),
and relatively
low spin temperatures, $T_{s}$ $<$ 1000 K, implied that
these absorbers were H I layers in which the gas
was cold and quiescent.

For these
reasons 
\citet{wolfeetal86}
 began a
survey for {\DLAs} by acquiring spectra of large numbers
of QSOs and then searching them for the presence of
damped {\lya} absorption lines. 
The survey for {\DLAs} 
had several advantages over surveys for 21 cm absorption
lines.
For example
the redshift interval
covered
by a single optical spectrum, $\Delta z$
$\approx$ 1, 
is large compared to 
that sampled by bandpasses then available 
for  21 cm surveys, 
$\Delta z$ $\approx$
0.02.
Second, optical spectra of QSOs are obtained toward
continuum sources with diameters less than 1 pc, whereas
the diameters of
the associated background radio sources typically
exceed 100 pc  at the low frequencies
of redshifted 21 cm lines. As a result the survey
was capable of detecting compact gaseous configurations 
with low surface covering factors that would have been
missed in 21 cm surveys. Another advantage of 
optical surveys is the large oscillator strength, $f_{21}$=0.418,
of the {\lya} transition (by comparison $f_{21}=2.5{\times}10^{-8}$ 
times a stimulated emission correction of 0.068K/$T_{s}$ for the 21 cm line),
which  allows for the detection of
warm H I, which  
is optically thin to 21 cm absorption
owing to high values of $T_{s}$
but 
optically thick in {\lya}. But this is
also  a disadvantage:
the strength of the {\lya} transition
combined with the high abundance of hydrogen
means that the more frequently occurring low column-density
clouds in which H is mainly ionized
will be optically thick in {\lya}. The result
is a profusion of {\lya} absorption lines, i.e., 
the {\lya} forest,  which dominate the absorption 
spectrum blueward of {\lya} emission
(see Figure~\ref{fig_lyaqso}). Although the {\lya} forest lines
act as excellent probes of the power spectrum and other
cosmological quantities 
\citep[see][]{tytleretal04a,mcdonald03},
they are potential sources
of confusion noise for the detection of damped
\lya\ lines, especially at $z$ $>$ 4, since the line
density per unit redshift increases with redshift. Identification
of damped \lya\ lines 
at $z$ $>$ 5.5 is essentially impossible because of \lya\ forest
confusion noise.

However, at $z \ <$ 5.5 the large column densities of H I in
galaxy disks or in any  other configuration
produce {\dla} absorption lines that are strong enough
to be distinguished from the {\lya} forest (see
Figure~\ref{fig_lyaqso}). 
Consider the equivalent
widths. 
At the time of the 
\citet{wolfeetal86}
survey the most accurate 21 cm maps of spiral
galaxies were obtained with the Westerbork radio interferometer.
These
showed the H I column densities  of galaxy
disks to decrease from {\nh} $\sim$ {\tenOO} at
their centers to {\nh}= {\NL} at
a limiting radius $R_{l}$=(1.5$\pm$0.5)$R_{26.5}$,
which was set by the sensitivity available with
Westerbork and comparable radio antennas. Here
the Holmberg radius, $R_{26.5}$, is the radius at
which the $B$ band surface brightness equals 26.5 mag arcsec$^{-2}$
\citep{bosma81}.
The rest-frame equivalent width of 
a {\dla} line created by an H I column density, {\nh}, is given by 
$W_{r} \approx 10 \times$[{\nh}/{\NL}]$^{1/2}${\AA}. Because the observed
equivalent width of a line formed at redshift $z$ is 
$W_{obs}$=(1$+ z$)$W_{r}$,
damped {\lya} systems with {\nh}${\ge}${\NL} will
appear in optical QSO spectra with
$W_{obs}  \ge 16$\AA\ for \DLAs\
redshifted redward of the
atmospheric cutoff (i.e.\ $z > 1.6$ for $\lambda_{atm} = 3200$\AA). 
Lines this strong 
are easily distinguishable from the $W_{obs}$
$\approx$ 3 {\AA} equivalent widths of
typical {\lya} forest lines. Furthermore, they
can be 
detected at low resolution and  moderate
signal-to-noise ratio. Since the goal of
the first survey for
{\DLAs} was to find absorbers with {\nh} $\ge$ {\NL},
a spectral resolution, {\Dlam}
=10 {\AA}, was sufficient for 
resolving candidate features.

\subsection{Modern Surveys and Identification of {\DLAs}}

Since the initial survey was published,
nine more surveys have been
completed for {\DLAs} with {\nh} $\ge$ {\NL}
(\citealt{lanzettaetal91}; \citealt*{lanzettawt95}; 
\citealt{wolfeetal95,storrielombardiw00,raot00,
ellisonetal01b,perouxetal03b,prochaskah04};Prochaska, Herbert-Fort,
\& Wolfe 2005).  
%
The identification of {\DLAs} is more complex than for other
classes of QSO absorbers. The {\lya} forest "clouds", 
which dominate the absorption spectrum blueward of {\lya} emission,
are abundant and easy to identify.
Similarly, surveys for C IV or Mg II absorption
systems rely on the detection of doublets with known wavelength
ratios, which are straightforward to locate redward of {\lya} emission.
By contrast, the task of surveys for {\DLAs} is to pick out a single
damped {\lya} line from the confusion noise generated by the
{\lya} forest. 
In particular, one must distinguish 
a single, strong {\lya} absorption line
created in high column-density gas with
low velocity dispersion, but broadened by radiation damping, from 
strong {\lya}
absorption features that are Doppler-broadened blends of several
lines arising from redshift systems with 
low column-density gas. The presence of narrow
{\lya} forest absorption
lines in the damping wings of the absorption profile is a further
complication which can distort the shape of the true line profile
in data of moderate or low signal-to-noise ratios 
(see Figure~\ref{fig_lyafits} for examples).

The most widely used
strategy for discovering {\DLAs} was first introduced 
by 
\citet{wolfeetal86}
and later refined by 
\citet{lanzettaetal91}
and 
\citet{wolfeetal95}.
First, a continuum is fitted to the entire QSO
spectrum blueward of {\lya} emission. Then {\dla} candidates are
identified as absorption features with rest equivalent widths $W_r$
exceeding $W_{thresh}=5$\AA\ . This
conservative criterion corresponds
to \ $\N{HI} \geq 5 \times 10^{19}$cm$^{-2}$,
which guarantees that few systems
with {\nh}  above the completeness limit of {\NL} will be missed. 
The search is carried out in the redshift interval
$z$=$[z_{min},z_{max}]$ 
where $z_{min}$ generally corresponds to
the shortest wavelength 
for which 
$\sigma(W_r) < 1$\AA\
and $z_{max}$ is set 3000 {\kms}
below $z_{em}$ to avoid contamination by the background QSO.
Finally, 
a Voigt profile is 
fitted to the {\lya} profile to determine the value of {\nh}.
Where possible, the centroid is identified from the redshift
determined by metal lines outside the \lya\ forest.
This is particularly important at $z>3$ where line-blending from
the \lya\ forest often contaminates the damping wings 
(e.g.\ Figure~\ref{fig_lyafits}).
The surveys were time consuming
because the signal-to-noise and resolution of the
spectra used to acquire {\DLA} candidates were usually inadequate
for fitting Voigt profiles to the data. 
Therefore, follow-up
spectroscopy at higher spectral
resolution and with longer integration
times was usually necessary.

Recently 
\citet{prochaskah04} and Prochaska, Herbert-Fort, \& Wolfe (2005)
have
streamlined this process in a survey based on a single set
of QSO spectra drawn from the Sloan SDSS archive 
\citep{abazajianetal03}.
Because of the high quality,
good spectral resolution ($R \sim 2000$) and extended
spectral coverage of the data, the authors could
fit accurate Voigt profiles to the same data used
to find {\DLA} candidates. 
The authors also bypass the time-consuming
step of fitting a continuum to the QSO spectrum blueward
of {\lya} emission
by searching for {\DLA} 
candidates in spectral regions with lower-than-average signal-to-noise
ratios; i.e., regions coinciding with broad absorption
troughs.   
The survey is not formally complete
to {\nh} = {\NL}, but the similarity between 
$d{\cal N}/dX$,  
the number of {\DLAs} encountered per unit absorption 
distance along the line of sight (see $\S$ \ref{sect:formalism}),   
in
their survey and previous surveys suggests they are more than
95$\%$ complete.  
The number of {\DLAs} for the SDSS DR2 and DR3 archives is
525. As a result the number of {\DLAs} in a statistically
complete sample now exceeds previous samples by an order
of magnitude at $z \ \sim$3 and several times at $z \ \sim$4.

While the  H I selection methods are successful at finding {\DLAs}
at $z$ $>$ 1.6, they have been unsuccessful at finding
large numbers of objects at lower redshifts.
This is partly due to the reduced interception probability per
unit redshift at low $z$ and partly because few QSOs have been
observed from space at UV wavelengths, which is required to detect
{\lya} 
at $z$ $<$ 1.6. 
To increase the number of low-redshift {\DLAs}
from the two confirmed objects detected in
previous H I selected surveys (see \citealt{lanzettawt95}),
 \citet{raot00} searched
for {\DLAs} in samples of QSO absorption systems selected for
Mg II $\lambda$$\lambda$ 2796.3, 2803.5 absorption.
Since Mg II absorption is present in every damped system in which it
could be observed, it turns out to be a reliable indicator
for the presence of {\dla}. Using this technique, Rao, Turnshek,
and collaborators have recently 
increased 
the sample size to  41 {\DLAs} with $z$ $<$ 1.6 (SM Rao, DA Turnshek, \&
DB Nestor 2004, priv. comm.).

The current sample of {\DLAs} that are drawn from 
surveys with statistically complete selection criteria comprises
over 600 redshift systems. While the
number of 
{\DLAs} is smaller than the $\approx$ 2350 objects comprising
the population of known Lyman Break galaxies (Steidel {\etal} 2003),
we expect the {\DLA} population to approach this number when all QSO spectra
from the Sloan database become available.

\subsection{The Significance of the N{\rm(H I)} $\ge$ {\NL} Survey Threshold}

The survey statistics cited above refer only to systems with
{\nh} $\ge$ {\NL}, which is an historical threshold set by
the H I properties of nearby spiral galaxies (see
$\S$ 1.1). Because the nature
of {\DLAs} is still not understood, their H I
properties may differ from those of nearby H I disks:
for example, CDM cosmogonies envisage 
{\DLAs} as merging protogalactic clumps 
\citep*{haehneltsr98}.
As a result,
it is reasonable to ask whether the {\NL} threshold is the
appropriate one. Indeed,
since the empirically determined 
frequency distribution of H I column densities 
increases with decreasing {\nh} (see Figure~\ref{fig_fNvsz}),
lower H I thresholds would be
advantageous because they would result in larger samples.

Fortuitously, the {\NL} threshold is optimal
for physical reasons unrelated to the properties of galaxy disks.
Rather, at large redshifts 
this  is the column density which divides neutral gas from 
ionized gas: at {\nh} $<$ {\NL} the gas is likely to be ionized
while at {\nh} $>$ {\NL} it is likely to be neutral.
The minimal source of ionization is background radiation due
to the integrated population of QSOs and galaxies. Using 
background intensities computed by 
\citet{haardtm96,haardtm03},
\citet{viegas95}
and  
\citet{prochaskaw96}
show that
the gas in most of the ``sub-{\dla}'' population 
(defined to have $10^{19} < ${\nh} $<$ {\NL}) described by 
\citet{perouxetal02,perouxetal03a}
is in fact significantly ionized with
temperature, $T$ $>$ 10$^{4}$ K.
This is a problem since
gas neutrality is a necessary condition if {\DLAs}
are to serve as  neutral gas reservoirs for star formation at high
redshift, a defining property of the population.
For this reason the  comoving density of H I 
comprising the sub-{\dla} population 
discussed by 
\citet{perouxetal03b}
should not be included in the census of gas
available for star formation. As a result, the sub-{\dla}
correction
to the comoving density of {\em neutral} gas, {\omgas},
should be ignored. 
We suggest that these ionization levels make ``super Lyman-limit system''
a more appropriate name for systems with 
$10^{19}<$ {\nh} $<$ {\NL}.


\section{THE NEUTRAL GAS CONTENT OF THE UNIVERSE}
\label{sect:neutralgas}

In this section we describe how the surveys allow us to measure
{\omgas}, the mass per unit comoving volume of neutral gas in
{\DLAs} at redshift $z$ divided by the critical density, $\rho_{crit}$.
The results, first derived by 
\citet{wolfe86} and 
\citet*{lanzettawt95} show that 
{\DLAs} 
contain most of 
the neutral gas in the Universe at redshifts $1.6<z<5.0$.

\subsection{Formalism}
\label{sect:formalism}

To estimate {\omgas} we first derive an expression for 
the column density distribution, {\fNX}.
Let the number of absorbers per sightline with H I column
densities and redshifts in the intervals ($N,N+dN$) and ($z,z+dz$)
be given by

\begin{equation}
  d{\cal N}(N,z) = n_{co}(N,z)A(N,z)(1+z)^{3}|cdt/dz|dNdz
\cmma
\label{eq:dN(n,z)}
\end{equation}

\noindent where $n_{co}(N,z)dN$ is the comoving density of absorbers
within ($N,N+dN$) at $z$ and $A(N,z)$ is the absorption cross section
at ($N,z$). Defining $dX{\equiv}(H_{0}/c)(1+z)^{3}|cdt/dz|dz$  
\citep{bahcallp69}
we have

\begin{equation}
  {d{\cal N}(X) \over dX}={\int_{N_{min}}^{N_{max}}dNf(X,N)}
\cmma
\end{equation}

\noindent where 

\begin{equation}
f(N,X){\equiv}(c/H_{0})n_{co}(N,X)A(N,X)
\cmma
\end{equation}

\noindent and where 
$N_{min}$ and
$N_{max}$ are minimum and maximum column densities\footnote {Note that
$dX/dz=(1+z)^{2}[{(1+z)^{2}(1+z{\Omega_{m}})}-{z(z+2){\Omega_{\Lambda}}}]^{-1/2}$}.
Therefore, 
one cannot infer the comoving density nor
the area of {\DLAs} from their incidence along the line of sight,
but only their product. 
Note, 
$d{\cal N}/dX$ will
be independent of redshift
if the product of the comoving
density and absorption cross section at ($N,X$) 
is independent of redshift. 
Since the gaseous mass per {\DLA} is given by
$\mu$$m_{H}$$N$$A(N,X)$, it follows from eq. (3)  that

\begin{equation}
\Omega_{g}={H_{0} \over c}{{{\mu}m_{H}} \over \rho_{crit}}\int_{N_{min}}^{N_{max}}dNNf(N,X)
\perd
\end{equation}

\noindent where $\mu$ is the mean molecular weight, which is included to
account for the contribution of He to the neutral gas content.

Using these expressions in the discrete limit, several authors have 
determined {\fNX} and its first two moments,
{\dndx} and {\omgas}, where

\begin{equation}
\Omega_{g}(X)={H_{0} \over c}{{{\mu}m_{H}} \over \rho_{crit}}{{\sum_{i=1}^{n}N_{i}} \over {\Delta X}}
\cmma
\label{eq:omegagas}
\end{equation}

\noindent and $n$ is the number of {\DLAs} within ($X, X+{\Delta}X$).
We now discuss each of these in turn.

\subsection{$f(N,X)$}
\label{sect:cdd}

Figure~\ref{fig_fNvsz} shows
the most recent determination of {\fNX} from the 
statistical sample  of over 600 {\DLAs} (Prochaska, Herbert-Fort,
\& Wolfe 2005). 
The figure also shows best-fit solutions for the three functional 
forms used to describe {\fNX}: a single power-law, {\fNX}
=${k_{1}}N^{\alpha_{1}}$;
a $\Gamma$ function (e.g. Pei \& Fall 1993) 
{\fNX}=${k_{2}}(N/N_{\gamma})^{\alpha_{2}}{\rm exp}(-N/N_{\gamma})$;
and a double power-law {\fNX}=${k_{3}}(N/N_{\gamma})^{\beta}$ where
$\beta$= $\alpha_{3}$ at $N \ < N_{d}$ and $\alpha_{4}$
at $N$ $\ge$ $N_{d}$. The single power-law solution with a best-fit
slope of $\alpha_{1}$ = $-$ 2.20$\pm$0.05 is a poor description of
the data since a KS test shows there is a 0.1$\%$ probability that the
data and power-law solution are drawn from the same parent population.
This result is in contrast with the {\lya} forest where a single power-law
with $\alpha_{1}$ $\approx$ $-$1.5 provides a good fit to the 
data (Kirkman \& Tytler 1997).

Although a single power-law is a poor fit to the observations, the
{\fNX} distribution is steeper than $N^{-2}$ at large column densities.
This is illustrated by the other two curves in Figure 3 that
show the $\Gamma$ function (dashed line) and the double power-law
(dashed-dot line). Both solutions are good fits to the data.
Furthermore, the solutions provide 
good agreement between the `break' column densities $N_{\gamma}$
and $N_{d}$, and between the power-law indices at low
column densities, which approach a `low-end' slope $\alpha$=$-$2.0.
Most importantly, both solutions indicate $\alpha  \ << \ -$2.0 at
$N \ \ge$ 10$^{21.5}$ cm$^{-2}$: the significance of this very 
steep slope at the `high end' will  be explored further in
$\S$ 2.4.

Prochaska, Herbert-Fort, \& Wolfe (2005)
 also find evidence for evolution in {\fNX},
which will be  clearly visible in the redshift dependence of {\dNdX}, the
zeroth moment of {\fNX}, and of {\omgas}, the first moment of
{\fNX} (see  $\S$ 2.3 and $\S$ 2.4). At low column densities, 
{\fNX} increases with redshift by a factor of 2 at $z$  $\ge$
2.2. Prochaska, Herbert-Fort, \& Wolfe (2005)
detect a similar evolution at higher
values of $N$. By contrast the shape of {\fNX} does not
appear to evolve with redshift. This is in disagreement with the
earlier results of \citet{storrielombardiw00} and \citet{perouxetal03b}
who used much smaller samples
to claim that {\fNX}  steepened at $z$ $>$ 3.5..

\subsection{$d{\cal N}/dX$}
\label{sect:dndx}

In Figure~\ref{fig_nx} we plot the most
recent evaluation of {\dndx} versus $z$ for {\DLAs} with
$z${$\ge$}0 (see Prochaska, Herbert-Fort, \&
Wolfe 2005). The solid line traces
the value of {\dNdX} derived in a series of 0.5 Gyr time
intervals to reveal the effects of binning.
The data points at $z$ = 0 are three estimates 
of ({\dndx})$_{z=0}$ based on the H I properties of
nearby galaxies.
The figure shows  a 
decrease by a factor of two in {\dNdX} from $z$=4 to 2. From equations 2 and 3
we see that the decrease in {\dNdX} reflects a decrease in either H I
cross section, comoving density, or of both quantities.
Prochaska, Herbert-Fort, \& Wolfe (2005)
use the Press-Schechter formalism to show that significant variations
in comoving density with time are unlikely to occur. Therefore,
within the context of CDM models the most likely explanation
for the changes in {\dNdX} is a decrease in H I cross section
with time. This 
is probably due to feedback mechanisms such as
galactic winds. 

Figure~\ref{fig_nx} also shows that {\dNdX} at $z$ $\approx$ 2 
is consistent with the present-day value; i.e., the data
are consistent with an unevolving population of galaxies
\footnote{
\citet*{ryanweberws03} 
present evidence that {\fNX} at $z$=0 is significantly
lower in amplitude than the results at higher redshifts, 
but this result is puzzling
since comparison between the resultant {\dndx} at $z$=0 with the higher
redshift data in Figure~\ref{fig_nx} reveals no evidence for evolution.}.
By comparison, 
\citet{wolfeetal86}
required more than a factor of 4 increase with redshift
in {\dndx}.
The discrepancy arises from the earlier use of an Einstein-deSitter
rather than the modern {\lcdm} cosmology to estimate
$\Delta X$ intervals, and from the lower values
of ({\dndx})$_{z=0}$ used in the earlier work.
This result implies that between $z$ = 1 and 2 
smaller galaxies merged to produce
bigger ones such that the product 
of comoving density and total H I cross-section
for {\nh} $\ge$ {\NL} is conserved.



\subsection{\omgas}
\label{sect:omgas}


Figure \ref{fig_omega} shows the most recent determination of {\omgas}.
From Equation (4) we see that {\omgas} is sensitive to the upper
limit $N_{max}$ unless $\alpha$ $<<$ $-$ 2. This led to large uncertainties
in {\omgas} in previous work because $\alpha$ was not measured with
sufficient accuracy to rule out $\alpha$ $\ge$ $-$ 2.
However, with the large sample of over 600 {\DLAs} Prochaska,
Herbert-Fort, \& Wolfe (2005) use several
tests to show that
{\omgas} converges. First, they compute $\alpha_{1}$
for a single power-law fit to {\fNX} by
increasing $N_{min}$ from {\NL}. Using the full
sample of {\DLAs} they find 
$\alpha_{1}$ decreases with increasing $N_{min}$ from $-$2.2 
at $N_{min}$={\NL} to less than $-$3 at $N_{min}$ $>$ {\tenOO}.
At the same time they find that $\alpha_{1}$ is insensitive to
variations in $N_{max}$. Second, they compute the sensitivity
of {\omgas} to $N_{max}$. Both the double power-law and $\Gamma$
function solutions converge to the value indicated by the data
(Equation 5). By contrast the single power-law solution does not
converge. This is the first evidence that {\omgas} converges
by $N$ $\approx$ 10$^{22}$ cm$^{-2}$.

Next consider the redshift evolution of {\omgas}. Starting
at the highest redshifts, no increase of {\omgas} with
decreasing $z$ is present at $z$ $>$ 3.5, contrary to earlier claims
(\citealt{storrielombardiw00}; \citealt{perouxetal03b}).
On the other hand,
Figure \ref{fig_omega}
shows the first statistically significant evidence that
{\omgas} evolves with redshift. Specifically, {\omgas} decreases
from 1{$\times$}10$^{-3}$ at $z$=3.5 to 0.5{$\times$}10$^{-3}$
at $z$=2.3, which mirrors the decline in {\dNdX} discussed
in $\S$ 2.3. The same mechanism is likely to cause the decline in
both quantities; i.e., a decrease in H I cross section due to
feedback. But at $z$ $<$ 2.3 the picture is somewhat confusing. 
Figure \ref{fig_omega} shows an {\em increase} of 
{\omgas}  by $z$ $\sim$ 2, which is consistent
with the values of {\omgas} in the two lower redshift bins at 
0 $<$ $z$ $<$ 2. Indeed, the data are consistent
with no evolution, if one ignores the redshift interval
centered at $z$ = 2.3. However, Prochaska, Herbert-Fort,
and Wolfe (2005) emphasize that 
the uncertainties in the data at 0 $<$ $z$ $<$ 2.3 are much
larger than at $z$ $>$ 2.3, and thus such conclusions 
should be treated with caution.
 
Next, we compare the high-$z$ values of {\omgas} with
various mass densities at $z$ = 0. First, comparison 
with the current density of visible stars, {\omstar}, reveals
that {\omgas} at $z$ $\approx$ 3.5 is a factor of 2 to 3 lower than
{\omstar}:
if the census of visible stars were restricted to stellar disks,
then {\omgas} at these redshifts would exceed $\Omega_{*}(z=0)$.
A straightforward interpretation of this concurrence is that
{\DLAs} provide the neutral gas reservoirs for star formation 
at high redshifts.
However, since
{$\Omega_{*}(z=0)$} exceeds
{$\Omega_{g}(z{\approx}3.5)$, the reservoir must be replenished
with new neutral gas before the present epoch. Further evidence for
replenishment is that it is required to compensate for gas
depletion due to star formation 
detected in {\DLAs} (see  $\S$~\ref{sect:sfr}). As a result the
``closed box'' hypothesis for evolution in {\DLAs} is unlikely
to be correct (see \citealt{lanzettawt95}). 

Figure \ref{fig_omega} also shows that {\omgas} at $z$ $\approx$ 3.5
is significantly higher than {\omgas} at $z$ = 0, which is deduced
from surveys for 21 cm emission.
Therefore, {\DLAs} provide direct evidence
for the widely held theoretical view 
that the neutral gas content of the Universe was larger at high redshifts 
than it is
today. Figure \ref{fig_omega} also shows that
{\omgas} at $z$ $\approx$ 3.5. is at least a factor of 10 greater
than {\omstar}
in dwarf irregular galaxies, which argues against the idea that
{\DLAs} evolve into such objects (e.g. \citealt{jimenezbm99}).

Therefore, since Lyman limit systems do not contribute
significantly to the {\em neutral} gas content at 
any redshift (see Prochaska, Herbert-Fort, \& Wolfe 2005)
and
 ignoring the possible existence of a significant
population of dusty giant molecular clouds, we conclude that  
{\DLAs} 
contain most
of the gas available for star formation at $z>1.6$.  
At low redshifts the ionizing background is reduced and lower {\nh} 
systems might be mainly neutral. But at $z=0$, 
{\citet{minchinetal03}} find a paucity of galaxies with column densities 
less than {\nh} = {\NL} measured from 21cm emission, implying that 
at the lowest redshifts, {\DLA} column densities comprise 
most of the neutral gas in the Universe. As a result, {\DLAs}
dominate the neutral-gas content of the Universe in the redshift
interval $z$ = [0,5].

Of course, all of these conclusions ignore 
obscuration by dust in {\DLAs}, which may have  biased the 
form of {\fNX} 
(see \citealt*{peifb91,fallp93}).  
We discuss this possibility
in $\S$~\ref{sect:dust}. They also ignore biasing due to
lensing, which may be present (see $\S$ 11).

\subsection{Model Comparisons}
\label{sect:models}

Here we discuss attempts to
use the H I content of
{\DLAs} to test models
of galaxy formation and evolution. 

\subsubsection{Passive Evolution:   The Null Hypothesis}
\label{sect:passive}

 The evolution of
H I content within the null hypothesis of passive evolution 
has been modeled by 
\citet*{boissierpp03}.
%
In this scenario
damped {\lya} absorption arises in disk galaxies with the
comoving density of current spiral galaxies. 
The models are 
hybrids of passive evolution, in which the
H I content is changed only by processes
of stellar evolution, and the spherical
collapse model in an expanding Universe, 
in which  high-$z$ disks
are smaller than current disks. The models are successful
in explaining the lack of evolution in {\dndx}, {\fNX}, and {\omgas}
at $z$ $<$ 2 but fall short at higher redshifts
because of delayed disk formation. For these
reasons, the  authors
suggest the added presence of a population of 
low-surface brightness, gas-rich galaxies at $z$ $>$ 2.
However, evidence against the ``closed box'' hypothesis discussed
in $\S$ 2.4 is a difficult challenge for this and all passive evolution 
models. 

\subsubsection{Numerical Simulations}
\label{sect:nums}

Several aspects of cosmology and galaxy formation have been
examined through comparisons of
numerical simulations of galaxy formation in a CDM Universe
with the observed H\,I properties of {\DLAs}. 
A critical feature of these models is 
that gas falling onto dark-matter halos
is heated to their virial temperatures, then
cools off, and collapses toward the central regions of
the halos.
Galaxies arise
from the formation of stars out of
the cool (presumably neutral) collapsed gas 
and evolve through mergers between dark-matter halos 
and further infall of gas onto the halos.

The first studies \citep{mab94,klypinetal95}
constrained the cosmological 
mass constituents through comparisons with {\omgas}.  The
observations severely restricted the contribution from a hot component
(i.e.\ neutrinos) as these cold+hot dark matter cosmogonies underpredicted
structure formation at early times
\citep{katzetal96}. Subsequent papers by 
\citet{gardneretal97a,gardneretal97b,gardneretal01}
examined the properties of {\DLAs}
in their smooth particle hydrodynamic (SPH) simulations.  These
models generically underpredicted the incidence of {\DLAs}, 
which the authors argued was due to an insufficient
mass resolution of 10$^{11}${\msolar}.
They did find reasonable agreement with the data, however, by
extrapolating to halos with $M_{h}$ $>$ 10$^{10}$ {\msolar}
using the Press-Schechter formalism, and by
assuming that the H\,I cross-section
followed the power-law expression $A \propto v_c^{1.6}$.
As stressed by \cite{prochaskaw01}, this power-law expression implies a $v_c$
distribution that is incompatible with the observed {\dla} velocity widths
(see $\S$~\ref{sect:kinematics}).

\citet*{nagaminesh04a}
recently analyzed 
a comprehensive set of high resolution SPH simulations of a {\lcdm} Universe.
%
By contrast with \citet{gardneretal97a}
they find that halo masses down to 
$M_{h}$ $\approx$ 10$^{8}$\,{\msolar} 
contribute to the H I cross-sections: halos 
with $M_{h} < 10^8$ {\msolar}
do not contribute since they contain only photo-ionized gas.
In turn, Nagamine {\etal} find a steeper power-law expression
$A$ $\propto$ $v_{c}^{2.7}$,  
i.e., massive halos make a larger contribution to {\dndx}
than in previous models.  
They conclude that \citet{gardneretal97a} predicted
an overabundance of {\DLAs} with $M_{h}$ $<$ 10$^{10}$
{\msolar} because the slope of their $A$ versus $v_{c}$
relation was too shallow.
\citet*{nagaminesh04a} were the first to
include mass loss of neutral gas due to winds,
which  increases the median halo-mass
contribution to {\dndx} to 10$^{11}$ {\msolar}. 
Winds also prevent an overabundance of
{\omgas} at $z$ $>$ 4, but 
underpredict {\omgas}
at $z$ $<$ 4. 
\citet{nagaminesh04a}
continue to find a deficit of
{\DLAs} with {\nh} $<$ {\tenOO}.
The deficit of systems with low
{\nh} is a generic effect seen in most
(e.g.\ Figure~3 in \citealt{katzetal96})
but not all 
numerical simulations 
\citep{cenetal03} 
and is a shortcoming that needs to be addressed.
On the other hand, the Nagamine, Springel, \& Hernquist (2004a)
models are the most successful in reproducing the evolution of
{\omgas) at $z$ $>$ 2 (see Figure 5).

\subsubsection{Semi-analytic and Analytic Models}
\label{sect:sams}


The semi-analytic models were proposed to 
include processes beneath the resolution of the
numerical simulations 
with phenomenological descriptions of 
star formation, gas cooling,
and the spatial distribution of the gas. 
The latter is included
since the simulations failed to reproduce the correct
sizes and angular momenta of present-day galactic
disks 
\citep{navarros00},
and we do not know whether the
simulations produce the correct spatial distribution of
neutral gas at $z$ $\sim$ 3. This is a concern because {\DLAs} are a
cross-section weighted population of high-redshift
layers of neutral gas and therefore the results will be sensitive
to the gas distribution at large impact parameters. 
By contrast with the numerical simulations, one uses analytic
expressions from Press-Schechter theory or its extensions
(\citealt{presss74};\citealt*{shethmt01})   
to compute the
mass function of halos that evolves from a given power spectrum.

\citet{mom94} 
modeled {\DLAs} with the Press-Schechter 
formalism in both mixed Cold+Hot dark 
matter  and {\lcdm} cosmologies. 
\citet{kauffmann96}
used a SCDM cosmology
([{\omgm},{\omgv},$h$]=1.0,0.0,0.5) to construct improved semi-analytic
models for {\DLAs}.
She assumed spherical geometry 
for the halos and let
the neutral gas in a given halo be confined to a
smaller,
centrifugally supported disk. To compute the
radial distribution of the neutral gas, the angular momentum
per unit mass of the disk was set equal to that of the halo. 
Using Monte Carlo
methods, she computed the formation 
and growth of individual halos with time. 
The neutral gas content  was assumed to be regulated by
accretion due to mergers 
and
star formation, but feedback due to winds was omitted. 
In common with
the 
\citet*{nagaminesh04a}
simulations, the 
\citet{kauffmann96}
models
(1) reproduced the {\omgas} relation, and (2) exhibited
a deficit of systems with {\nh} $<$ {\tenOO}.

\citet*{momw98}
 constructed models for disk
formation that were also based on the Press-Schechter formalism.
These models extended the work of 
\citet{kauffmann96}
 by considering disks drawn from a distribution of halo spin
parameters, $\lambda_{H}$, rather than using Kauffmann's technique
of assigning the mean value of $\lambda_{H}$ 
to each disk. Since disks detected in a survey
for {\DLAs} are drawn from a cross-section
weighted sample  favoring
bigger disks, the  
distribution of $\lambda_{H}$ 
will be skewed to values higher than the unweighted
mean. The result is
larger H I cross-sections and 
higher detected rotation speeds. 
Consequently, they found agreement between the
predicted and observed {\dndx} relation. 
\citet{malleretal01}
 then suggested a model
in which the gas is in extended
\citet{mestel63}
 disks in which
the surface density falls off inversely with radius. In this
case the disks overlap and as a result the observed 
{\fNX} is reproduced; i.e., there is
no deficit of systems with {\nh} $<$ {\tenOO}. However, 
it is unclear whether such
extended disks will either form or survive sufficiently
long to contribute to the H I cross-sections of {\DLAs}.
Furthermore, most semianalytic models overestimate {\omgas}
at $z$ $>$ 2, since they underpredict feedback processes at
these redshifts (see Figure 5).

\section{CHEMICAL ABUNDANCES}
\label{sect:chemicalabund}

Because the 
{\DLAs} comprise the neutral gas reservoir for star formation
at high redshifts, a determination of their metal content is
a crucial step  for understanding the chemical
evolution of galaxies.  
Therefore, the mass of metals
per unit comoving volume that they contribute
indicates the level to which the
neutral gas reservoir has been chemically enriched. 
Since the metal abundances of {\DLAs} have been
determined in the redshift
interval $z$ = [0, 5], it is now possible to track the chemical
evolution of the reservoir back $\approx$ 10 Gyr to the time the
thin disk of the Galaxy formed ($z$ = 1.8 
for the WMAP cosmological parameters 
adopted here), and to earlier epochs.
As a result, one can
construct an ``age-metallicity'' relation not just for the
solar neighborhood 
\citep[see][]{edvardssonetal93}
 but for a
fair sample of galaxies in the Universe.

In this section we describe the main results that have emerged
from abundance studies of {\DLAs}. This subject has recently been
reviewed in an excellent article by 
\citet{pettini04}.

\subsection{Methodology}
\label{sect:mtlmeth}

The element abundances of the {\DLAs} are the most accurate measurements
of chemical enrichment of gas in the high-redshift Universe.
The measurements are accurate for
several reasons: (1) For the majority of {\DLAs}, hydrogen is mostly
neutral, i.e., H$^{0}$/H=1, and most of the abundant elements are
singly ionized, though a minority are neutral,  i.e., 
Fe$^{+}$/Fe=Si$^{+}$/Si=1, etc., while O$^{0}$/O=N$^{0}/$N=1.  
The singly ionized elements have 
ionization potentials of their neutral states that
are lower than the ionization potential of hydrogen, 
IP(H) (=13.6 eV). 
With Lyman limit optical depths, $\tau_{LL} \ >>$ 10$^{3}$,
{\DLAs} are optically thick at
photon energies,  IP(H) $\le$ $h{\nu}$ $\le$ 
400 eV. 
As a result, only photons with $h{\nu}$ $\le$ IP(H) and
$h{\nu}$ $\ge$ 400 eV penetrate deep into the neutral gas. 
When FUV photons with $h{\nu}$ $\le$ IP(H) penetrate, 
they photoionize the neutral state of each element to the  
singly ionized state. But this state is shielded from 
photons with IP(H) $\le$ $h{\nu}$ $\le$ 400 eV,
which would  otherwise photoionize  the elements to higher states.
Photons with $h{\nu}$ $>$ 400 eV
will produce species that are doubly ionized (e.g.\ Fe$^{++}$ and Al$^{++}$)
and singly ionized (e.g.\ Ar$^{+}$),
but because the photoionization 
cross-sections are low at such high photon
energies, the ionization rates are low. 
It is possible
to detect all of these species because they exhibit resonance transitions
that are redshifted to optical wavelengths accessible
with ground based spectrographs. 
(2) In $\S$ 1 we saw that Voigt fits to the damped
{\lya} profiles result in typical errors of 0.1 dex
in {\nh}. As we shall see, the errors in the column
densities which give rise to the narrow low-ion lines are typically
0.05 dex. Consequently, errors in [X/H]
\footnote{ Here and  in what follows the relative abundance
of elements X and Y is defined with respect to the solar
abundance on a logarithmic scale; i.e. [X/Y]=
log$_{10}$(X/Y)$-$log$_{10}$(X/Y)$_{\odot}$.}
are relatively
low (typically
about  0.1 dex). 
(3) Column densities are 
straightforward to measure from resonance lines, since their optical depths 
are independent
of poorly determined physical parameters such as the density and temperature
of the absorbing gas.

By contrast, 
abundance determinations for the
other constituents of the high-redshift
Universe are more uncertain primarily because 
the gas is ionized. As a result, the abundances are subject
to ionization corrections
which depend on uncertainties in the shape of the ionizing continuum
radiation and on the transport of such radiation. Furthermore,
the strengths of QSO emission lines
depend on the temperature
and density of the emitting gas as well as 
uncertain photon escape probabilities in
the case of resonance scattering.
Typical error estimates are about 50 $\%$ per object, which
is several times higher than for {\DLAs}
(see \citealt{hamannf99}
 for an excellent review of this subject).

Figure~\ref{fig_q1108} 
shows examples of absorption profiles
obtainable with the HIRES  Echelle spectrograph mounted on the Keck\,I 10-m
telescope. The figure shows velocity profiles for abundant low
ions in two {\DLAs}.
As in most {\DLAs} the gas that gives rise to low-ion absorption
lines in these two objects is comprised of 
multiple discrete velocity structures of enhanced density;
i.e., clouds. To infer the ionic column densities 
required for element abundance determinations, one integrates
the ``apparent optical depth'' 
\citep{savages91,jenkins96} over the velocity profile.

To illustrate the essentials of abundance determinations we focus on
two {\DLAs}, one metal-poor
(DLA1108$-$07
at  $z$ = 3.608) 
and the other metal-rich 
(DLA0812$+$32 
at $z$ = 2.626)
\footnote{Here and in what follows we designate a {\DLA}
toward a QSO with coordinates hhmm{$\pm$}deg as
DLAhhmm$\pm$deg}. 
The corresponding velocity profiles
in Figure~\ref{fig_q1108} 
describe the challenges as well as the
advantages of measuring \dla\ abundances.
First consider the challenges.
The abundance of carbon has not been accurately determined for
any {\DLA} because the
only resonance transition outside the {\lya} forest, 
C II $\lambda$ 1334.5, 
is not only saturated in the metal-rich system (Figure~\ref{fig_q1108}b), 
but is also
saturated in the metal-poor system
(Figure~\ref{fig_q1108}a).  
The availability
of several O I transitions makes it possible to
place bounds on the oxygen abundance. 
Several authors used
saturated O I $\lambda$ 1302.1 
to obtain lower limits and the weaker
O I $\lambda$ 971.1 or O I $\lambda$ 950.8
transitions
for upper limits since the
latter transitions are usually blended with {\lya} forest
absorption features 
(\citealt{dodoricom04,molaroetal00}; \citealt*{dessaugeszavadskypd02}).
On the other hand, 
a direct determination of [O/H] is possible for
the metal-rich system shown in Figure~\ref{fig_q1108}b  because this is
the only known case in which an unsaturated transition, 
O I] $\lambda$ 1355.6, is detected. 

Second consider the advantages. Abundance
determinations are possible for Fe, Si, and S because of
the presence of transitions with a wide range of oscillator
strengths. In the case of
Fe, the oscillator strengths $f_{1611.2}$ = 0.00136  
for Fe II $\lambda$ 1611.2 and
$f_{1608.4}$ = 0.0580   for Fe II $\lambda$ 1608.4. Thus, in the
metal-poor {\DLA} in Figure~\ref{fig_q1108}a
Fe II $\lambda$ 1611.2 is undetected, while unsaturated Fe II $\lambda$ 1608.4 
is detected.
By contrast, in the metal-rich {\DLA} in Figure~\ref{fig_q1108}b
unsaturated Fe II $\lambda$ 1611.2 is detected, 
while
Fe II $\lambda$ 1608.4 is saturated. In both
systems the iron abundance is determined from the unsaturated transitions.
Similarly, Figure \ref{fig_q1108} also demonstrates how the 
Si II $\lambda$ 1304.3, 1808.0 pair of transitions determines
the silicon abundance for the two {\DLAs}.


\subsection{Metallicity}
\label{sect:metallicity}

Whereas the abundance ratios discussed above refer only to
elements in the gas phase, some fraction of each element could
be depleted onto dust grains, as in the Galaxy ISM
\citep{jenkins87}.
This 
possibility was recognized early on by
\citet*{meyerwy89}
 and 
\citet*{pettinibh90}
 who made use of the Zn II $\lambda \lambda$ 2026.1, 2062.6
doublet to measure the metallicities of {\DLAs}.
Zn is well suited for this purpose because
Zn is relatively undepleted in the ISM with a mean
depletion of [Zn/H] $\approx$ $-$0.23 \citep{savages96}.
Moreover
Zn was believed to be an 
accurate tracer of Fe peak elements since [Zn/Fe] $\approx$
0 for stars with metallicities, $-$2.0 $<$ [Fe/H] $<$ 0
(but see discussion below). In addition
the
combination of the low solar
abundance of Zn
and the oscillator strengths 
of the Zn II
transitions implies they should be unsaturated
for  {\nh} $\le$ {\tenOO}, 
provided the velocity dispersion of 
the gas, 
$\sigma_{v} \ \ge$ 4 \ {\kms}. 
Because of its proximity in wavelength,
the 
Cr II $\lambda \lambda$ 2056.2, 2062.2, 2066.1 triplet was used
to study depletion, since
most of the Cr in the ISM is locked up in grains (Jenkins 1987).
%


In subsequent surveys
on several 4 m class telescopes, 
Pettini and colleagues 
\citep{pettinietal94,pettinietal97b,pettinietal99}
increased the size of their sample
%
and confirmed that {\DLAs} are
metal-poor  in the redshift interval $z$ = [0.5, 3.0]. 
\citet{pettini04}
found that
the cosmic metallicity $<Z>$ = $-$1.11$\pm$0.38, where $<Z>$ is defined as the
log of the ratio of the comoving densities 
of metals and gas,  $\Omega_{metals}$/{$\Omega_{g}$},
relative to the solar abundance; i.e. from Equation~\ref{eq:omegagas}

\begin{equation}
<Z>={\rm log}_{10}{\Biggl[} {{\sum_{i=1}^{n}10^{[{\rm M/H}]_{i}}N_{i}} \over {\sum_{i=1}^{n}N_{i}}}{\Biggr]}-{\rm log}_{10}{\rm (M/H)}_{\odot}
\cmma
\end{equation}

\noindent where M stands for the metallicity indicator, which  in this
case is Zn.
Second, 
surprisingly, there is no positive evidence
for redshift evolution. Specifically, 
\citet{pettini04}
finds no statistically significant
evidence for redshift evolution in
$<Z>$. 
This is contrary  
to  most models of chemical evolution (see $\S$~\ref{sect:chemev}), which predict
an increase in 
the mean Zn abundance 
with decreasing redshift, and further predict that the metallicity
should approach  $<Z> = 0$,
by the current epoch.
The sub-solar values of $<Z>$ at $z<1$ 
raised the possibility that {\DLAs} do not  
evolve into normal current galaxies 
\citep{pettinietal99}.  
%

Further progress was achieved with
the completion of a larger survey of over 120
{\DLAs} carried out primarily on the Keck 10-m telescopes
\citep{prochaskaetal03b}. 
In this survey most of the metallicities,
[M/H], are obtained from measurements of $\alpha$-enhanced elements 
Si, S, and O in order of decreasing priority and in
a few cases from Zn. 
Like Zn, S and O are volatile elements which are essentially undepleted
in  the ISM. While the refractory element Si is depleted in the
ISM, it is only mildly depleted in  {\DLAs}, where
Si tracks S, i.e., 
[Si/S] $>$ $-$0.1 
\citep{prochaskaw02},
 and thus can generally be used as an unbiased metallicity tracer.
Furthermore,
since S and Si have higher solar abundances than Zn, they
can be used to probe down to metallicities below the
Zn threshold of [Zn/H]$\approx -$1.7. In addition, the shorter wavelengths
of crucial transitions  such as S II $\lambda$ 1250.5
and Si II $\lambda$ 1304.3 allow one to
obtain metal abundances at higher redshifts
than are accessible with the Zn II transitions alone.
Note, the idea of 
combining abundances of Zn and
$\alpha$-enhanced elements is plausible
if Zn is a tracer of elements such as S, Si, etc.
The recent finding by
\citet{prochaskaw02}
that
[Si/Zn] = 0.03$\pm$0.05 supports this hypothesis.
Further support comes from the finding that
[Zn/Fe] ranges between 0.10  
to 0.20 
\citep{prochaskaetal00,nissenetal04,bihainetal04}
in stars with [Zn/H]~$> -1.5$, which indicates Zn is not
a strict tracer of Fe peak elements. In fact there is
currently little reason to expect [Zn/Fe]=0 aside from 
a coincidence related to the
star formation history of the Galaxy 
\citep*{fennerpg04}.


The results of 
\citet{prochaskaetal03b}
 are shown in
Figure \ref{fig_cheme} (updated to include new data at $z<1.5$ from
\citealt{kulkarnietal05} and \citealt{raoetal05}). The new survey confirms the
low metallicities of {\DLAs} 
found by Pettini and colleagues.  However, the greater
accuracy and larger redshift range 
of the new survey allows one to draw additional conclusions.
First, there are no {\DLAs} with [M/H] $<$ $-$2.6. This limit
is robust because there are no {\DLAs} without significant
metal absorption. 
Second, \citet{prochaskaetal03b}
find statistically significant
evidence for a linear increase of
 $<Z>$ with decreasing $z$. 
This result is robust owing to the large value
of $\sum_{i}^{m}N_{i}$. This is important since the shape
of {\fNX} indicates that $<Z>$ is sensitive to the
metallicity of systems with the largest values of {\nh}.
Because $\sum_{i}^{m}N_{i}$ $>1 {\times} 10^{22}$ cm$^{-2}$
in each of the high-redshift bins, 
only unusual, very metal-rich systems with {\nh} $>$ 10$^{22}$ 
cm$^{-2}$ could increase $<Z>$ significantly; i.e.,
only systems which depart significantly from the current 
{\nh} versus [M/H] relation could cause a marked
increase in $<Z>$.
Earlier claims for
evolution had statistical significance
lower than 3$\sigma$ and sampled
lower values of $\sum_{i}^{n}N_{i}$
\citep{kulkarnif02,vladilo02b}.
 


The ``age-metallicity'' relationship depicted in Figure~\ref{fig_cheme} 
provides new information about the enrichment history of {\DLAs}.
Specifically, the absence of
any system with a metallicity [M/H] $<$ $-$ 2.6
sets the {\DLAs} apart from the {\lya} forest.
From their analysis of the {\lya} forest, 
\citet*{simcoesr04}
 find a median abundance,
[C,O/H] = $-$2.8 and find that 30 $\%$ of their sample
have
[C,O/H] $<$ $-$3.5. 
\citet{schayeetal03}
find similar results for [C/H]. While they
deduce a higher median abundance for Si, i.e.,
[Si/H] = $-$2.0, about 40 $\%$ of their systems
are predicted to have [Si/H]  $<$$-$2.6. Clearly
the bulk of the {\dla} population has a different
enrichment history than the {\lya} forest. 
To explain the presence of the metallicity floor,
\citet{qianw03}
 use a standard
chemical evolution model to show that star formation
in {\DLAs}
results in a rise in metal
abundance which is so rapid that the probability for
detecting systems with [M/H] $< \ -$ 2.6 is exceedingly small.


Figure~\ref{fig_cheme}
also poses several dilemmas  for models of
chemical evolution.
First, if most of the gas in {\DLAs}
in the redshift interval
$z$ = [1.6, 4.5] 
were converted into stars,
then most of the stellar mass in
current galaxies would be metal poor, contrary
to observations \cite{tremontietal04}.
Second, 
the age-metallicity relation of the thin disk
of the Galaxy 
\citep{edvardssonetal93},
indicates that the thin
disk formed at lookback times less than
10 Gyr (i.e., $z$ $\approx$ 1.8) and that chemical
enrichment proceeded such that {\em all} thin disk stars formed
with [M/H] $> \ -1.0$. 
But the lower panel in Figure~\ref{fig_cheme} 
shows that [M/H] $< \ -1.0$ in about half of the {\DLAs}
with look-back times under 10 Gyr. While 
this result is
subject to the uncertainties of small number statistics and
observational bias,
the current metallicity trends in  low-redshift {\DLAs}
suggest that
{\DLAs} may not 
trace the star formation history of normal galaxies 
\citep{pettinietal99}.
%
%
%
Third,
if the linear increase of $<Z>$ with decreasing
redshift deduced at $z$ $>$ 1.6 is extrapolated
to $z$ = 0, the current mean metal abundance of galaxies
would be equal to $-$0.69 which appears too low.  But 
since the age-metallicity relationship
is essentially unconstrained by the data at $z$ $<$ 1.6,
such extrapolations should be treated with caution.
Indeed, $<Z>$ 
is doubling every Gyr at $z$ $>$ 2, and if we assumed $<Z>$
to be a linear function of {\em time} 
rather than redshift, then 
we would find that $<Z>$$\approx$0 
by $z \approx 0.5$. 

  
\subsection{Relative Abundances}



In $\S$~\ref{sect:metallicity}
we described evidence that {\DLAs}
are metal-poor. We discussed 
measurements of Zn and
Cr which indicate a gas-phase abundance ratio, [Zn/Cr] $>$ 0,
implying 
depletion of Cr by dust.
Since metal-poor stars in the Galaxy exhibit different
nucleosynthetic abundance ratios than the sun
\citep*{wheelerst89},
the abundance patterns observed in {\DLAs}
are probably due to some combination of  nucleosynthetic 
and dust depletion patterns. In this section we briefly
describe efforts to unravel these effects.
The reader is referred to a series of recent papers for a
more thorough discussion of these issues 
\citep{prochaskaw02,vladilo02b,pettini04}.  

\subsubsection{Depletion}
\label{sect:depletion}

The discussions of metallicity in the previous subsections
implicitly assumed that deviations of (X/H)$_{gas}$, the gas-phase abundance
of element X, from the solar abundance,
(X/H)$_{\odot}$, were only due to changes in the intrinsic abundance,
(X/H)$_{int}$. 
However, as mentioned
previously, (X/H)$_{gas}$ will also deviate from (X/H)$_{\odot}$ if
element X is depleted onto grains.  
One of the major challenges
in damped {\lya} research is to untangle these two effects.

The traditional method used by most workers in the field is
to compare the abundance of refractory element, X, to  volatile
element, Y, for which (X/Y)$_{int}$ = (X/Y)$_{\odot}$ 
in stars with a wide range 
of absolute abundances. 
In that case the condition [X/Y] $\ne$ 0 
is unlikely to have a nucleosynthetic origin. Rather
it likely arises from 
depletion of the refactory element onto grains.
Such a comparison is made in 
Figure~\ref{fig_ZnovFe}a,
which 
is a plot of [Zn/Fe] versus [Zn/H] for a sample
of 32 {\DLAs}. 
The figure reveals an unambiguous correlation between
[Zn/Fe] and [Zn/H]: a Kendall $\tau$ test rules out
the null hypothesis of no correlation at more than 99.7 $\%$ confidence. 
Because [Zn/Fe] $<$ 0.2 for Galactic stars with
[Fe/H] $>$ $-$2.0 , the most  plausible explanation
for this correlation
is that in {\DLAs}
the depletion of Fe onto grains increases with metal
abundance.
This argument also suggests that the depletion level decreases with
decreasing metal abundance. In that case [Zn/Fe] should approach
the intrinsic nucleosynthetic ratio, [Zn/Fe]$_{int}$, in the
limit [Zn/H] $<<$ 0. Determination of 
[Zn/Fe]$_{int}$ is important as it indicates the nucleosynthetic
history of these elements 
\citep{hoffmanetal96}, 
and it
is required for determining the dust-to-gas ratio, $\kappa$. 
For example, 
\citet*{wolfepg03} 
show that

\begin{equation}
{\kappa} = 10^{[{\rm Y/H}]_{int}}{\Biggl (}10^{[{\rm X/Y}]_{int}}-10^{[{\rm X/Y}]_{gas}}{\Biggr )} 
\cmma
\label{eq:kappa}
\end{equation}
where in this case X=Fe and Y=Zn.

Our discussion emphasizes the importance of estimating
the intrinsic, nucleosynthetic ratio, [Zn/Fe]$_{int}$. 
On the other hand, the observed Zn abundances are not sufficiently
low for the asymptotic approach to [Zn/Fe]$_{int}$ to be detected.
Specifically, because the Zn II transitions
are weak, only two
{\DLAs} with  [Zn/H] $< \ -$1.5 have been detected
(\citealt*{lusb98}; \citealt{molaroetal00,prochaskaw01}).
By contrast, clouds of such
low metallicity can be easily detected in the strong
Si II
transitions, as shown in Figure~\ref{fig_ZnovFe}b, which  plots
[Si/Fe] versus [Si/H]
down to [Si/H] = $-$2.6. The figure gives convincing evidence
that in the limit of vanishing
metallicity, [Si/Fe] approaches 
$\approx$ 0.3 rather than 0.
Furthermore,
at metallicities [Si/H] $>$  $-$1 we see 
evidence for an increase in [Si/Fe] with increasing
[Si/H]. This is the same phenomenon
seen in the [Zn/Fe] versus [Zn/H] diagram, which we plausibly attributed
to dust. The amplitude of the increase is weaker for
[Si/Fe]
because Si is weakly depleted.
On the other hand the increase of [Si/Fe] with [Si/H] is 
stronger evidence for dust since the nucleosynthetic
origin of Si is better understood than that of Zn \citep{hoffmanetal96}). 
\subsubsection{Nucleosynthetic Abundance Patterns}
\label{subsect:nucleosyn}

{{$\bullet$} $\alpha$ {\em Enhancements?}}

In $\S$~\ref{sect:depletion} 
we argued that the asymptotic behavior exhibited by
the [Si/Fe] ratio in the limit [Si/H] $<<$ 0 (Figure~\ref{fig_ZnovFe}b)
indicated a nucleosynthetic
ratio, [Si/Fe]$_{int}$ $\approx$ 0.3. This asymptotic limit is robust, 
as it is based on 
a large number (56) of precision measurements  obtained with 
echelle spectrometers on 8- to 10-m class telescopes. 
It also has important implications
for the chemical evolution of {\DLAs} if it equals the
intrinsic nucleosynthetic ratio. The reason is that
disk stars in the Galaxy exhibit a
systematic decrease of [$\alpha$/Fe] with increasing [Fe/H],
which indicates the increase with time of Fe contributed 
to the 
Galaxy ISM by type Ia  supernovae
relative to type II supernovae
\citep{edvardssonetal93}.
The presence of such trends 
in {\DLAs} would support the argument that they are the progenitors
of ordinary galaxies.
However, the existence of intrinsic $\alpha$ enhancements in {\DLAs}
is controversial. Using  the 
\citet{vladilo98,vladilo02a} models,
\citet{vladilo02b} 
and 
\citet*{ledouxbp02}
examined [Si/Fe] ratios corrected
for depletion effects and found median values of [Si/Fe]$_{int}$
consistent with solar.
Similarly, several studies of the depletion-free 
[O/Zn] and [S/Zn] ratios resulted in [$\alpha$/Zn]
ratios below those of metal-poor stars 
\citep{molaroetal00,centurionetal00,nissenetal04}.
%

Is it possible to resolve these conflicts? The lower values of
[$\alpha$/Zn] are compatible with the higher value of [$\alpha$/Fe]$_{int}$
indicated by Figure~\ref{fig_ZnovFe}b since 
[$\alpha$/Fe]=[$\alpha$/Zn] \\ $+$[Zn/Fe] and
\citet{prochaskaetal00}
and 
\citet*{chenkr05}
find [Zn/Fe]$\approx$0.15 for thick
disk stars with $-$0.9 $<$ [Fe/H] $<$ $-$0.6, while
\citet{nissenetal04}
 find [Zn/Fe] $\approx$ 0.1 for stars
with [Fe/H] $<$$-$1.8. Both results are consistent with
[Si/Fe]$_{int}$$\approx$ 0.2 to 0.4. 
If such $\alpha$ enhancements
are confirmed in {\DLAs}, one would conclude that the depletion 
corrections used by 
\citet{vladilo02a}
 and 
\citet*{ledouxbp02}
were too large. The latter are compatible with the dust content
suggested by the 
\citet*{peifb91}
 study of reddening in
{\DLAs}. But, since the more recent study of 
\citet{murphyl04}
argues against such a high dust content, the depletions 
used to correct the [Si/Fe] ratios may be too large.
We also note that [$\alpha$/Fe]$_{int}$=0
would imply significant depletion of Fe
at [Si/H] $< \ -$1, which is apparently at odds with the
insensitivity of [Si/Fe] to increases in [Si/H]
shown in  Figure~\ref{fig_ZnovFe}b. But this behavior may
result from two compensating effects: an increase in
[Si/Fe] due to Fe depletion and a decrease in [Si/Fe] due
to increasing Fe enrichment from type Ia supernovae.
Because of these uncertainties, it 
may be 
premature to use the [$\alpha$/Fe]
ratios in {\DLAs} as discriminants between
competing galaxy formation scenarios
\citep[e.g.][]{tolstoyetal03,vennetal04}.

{{$\bullet$} {\em Nitrogen Enrichment}}

\citet*{pettinilh95}
 first detected
nitrogen in {\DLAs} and suggested the [N/$\alpha$] versus
[$\alpha$/H] plane could be used as a clock to infer
their ages. According to 
\citet*{henryek00}
 a burst
of star formation would coincide with 
the injection of $\alpha$ elements into the surrounding
ISM by type II supernovae, followed by
the injection of $^{14}$N by AGB stars more than 0.25~Gyr later.
In the local Universe, one identifies a plateau of
[N/$\alpha$] values (with value $\approx -0.7$\,dex)
at low metallicity presumably consisting
of objects with ages greater than 0.25\,Gyr.
Within this interpretation,
metal-poor objects with ages less than 0.25\,Gyr would have
systematically lower
[N/$\alpha$] values
while more evolved {\DLAs}, in which $^{14}$N production
has caught up, would have [N/$\alpha$]$\approx$$-$0.7. 
Recent studies \citep{prochaskaetal02c,pettinietal02,centurionetal03}
have shown that the majority of [N/$\alpha$] values for
the {\DLAs} are near the plateau but there is a population
of {\DLAs} with 
[N/$\alpha$] $\approx$ $-$1.5 and very few {\DLAs} with intermediate
[N/$\alpha$] values.
The {\DLAs} with [N/$\alpha$]=$-$0.7
must be older than 0.25\,Gyr, indicating they are not
transient objects as suggested in some
schemes \citep[see][]{qianw03} but rather have ages 
comparable to the age of the Universe at $z \sim 3$.

The observations also pose a challenge to interpreting
the [N/$\alpha$] value as a strict age diagnostic.
If the ages of the {\DLAs} are comparable to 2.5\,Gyr, the
age of the Universe at $z$ $\approx$ 2.5, then 
fewer than 10$\%$ of the objects
would have [N/$\alpha$]=$-$1.5, contrary to current observation.
\cite{prochaskaetal02c}
interpret
the paucity of systems  with $-$1.5 $<$ [N/$\alpha$]
$<$ $-$0.7 as evidence for a bimodal IMF where systems near the
plateau at [N/$\alpha$]=$-$1.5 are drawn from an IMF truncated
from below at $M$=7.5 $M_{\odot}$.
In this scenario, {\DLAs}
near the plateau at [N/$\alpha$]=$-$1.5 need not be younger
than 0.25 Gyr, while systems
near the plateau at [N/$\alpha$]=$-$0.7 are objects
older than 0.25 Gyr
in which N production is due to 
the full mass range of intermediate-mass stars
drawn from a standard IMF. 
More recently 
\cite{molaro03} argued against
a bimodal IMF by  suggesting 
that {\DLAs} near the [N/$\alpha$]=$-$1.5 plateau are younger than
the 0.25 Gyr ``catch-up'' time. 
\citet{meynetm02}
and
\citet*{chiappinimm03}
suggested a mechanism for obtaining
a more uniform distribution between the
two plateaus.
They showed
that stellar rotation causes enhanced
mixing between the H-burning and He-burning layers, 
thereby producing
greatly enhanced
$^{14}$N production in massive stars.  
\citet{meynetm02}
reproduced the [N/$\alpha$]=$-$1.5 plateau for
stars with$M$=8-120 {\msolar} for rotation speeds
$v$sin$i$=400 {\kms}. Moreover, 
rotation may extend the effective lag time 
between N and $\alpha$ production for intermediate
mass stars beyond 0.25 Gyr.
However, the lack of many {\DLAs} with [N/$\alpha$]
in between $-1.5$ and $-0.7$\,dex
argues against this mechanism and in favor of the bimodal IMF.


{{$\bullet$} {\em Metal-Strong {\DLA}}

There exists a small subset of {\DLAs} for which the product of H\,I column
density and metallicity imply very strong metal-line transitions.
These `metal-strong' {\DLAs} yield abundance measurements for over
20 elements including O, B, Ge, Cu, and Sn.
Figure~\ref{fig_oddeven} shows the elemental abundance pattern
obtained for DLA0812+32, the $z$ = 2.626, metal-rich {\DLA} 
discussed in $\S$~\ref{sect:mtlmeth}  \citep*{prochaskahw03}.
This is the first {\DLA} for which absolute
abundances for B and Ge have been 
determined, and it is one of the few 
objects for which an accurate  measurement  of [O/H] is possible.
With the detection of over 20 elements, the metal-strong {\DLAs}
permit  a global inspection of its enrichment history.
The dotted line in Figure~\ref{fig_oddeven} is the solar abundance pattern 
scaled to oxygen. The good match to the data
shows that this system exhibits an enrichment pattern resembling 
that of the Sun. 
Furthermore, specific abundance ratios constrain various
nucleosynthetic processes in the young Universe.
For example, the abundance of the
the odd-$Z$ elements P, Ga, and Mn compared
to Si, Ge, and Fe, indicates an enhanced "odd-even effect" and
impact theories of explosive nucleosynthesis.
Similarly, measurements of the B/O ratio help develop theories of
light element nucleosynthesis while constraints on Sn, Kr, and
other heavy elements will test scenarios of the r and s-process.

\section{IONIZED GAS}


The best evidence for ionized gas in {\DLAs}
comes from the detection of C\,IV $\lambda$$\lambda$ 1548.1,
1550.7 in every object for which accurate spectra have
been obtained. Examples of the C IV velocity profiles are 
shown in \citet{wolfep00a} who compare them  
with the corresponding
low-ion profiles. The absence of a one-to-one 
alignment between the velocity components indicates
that the gas producing  C\,IV absorption is not
the same gas that produces low-ion absorption. 
The difference is not surprising. While the X-ray
background at $z$ $\sim$ 3 
is about 30 times brighter than at $z$ = 0 
\citep{haardtm03},
it 
is still not sufficiently bright
to produce C$^{3+}$ ions by photoionization in neutral gas with 
Lyman limit optical depths, $\tau_{LL}$ $>>$ 10$^{3}$
\citep{wolfeetal04}.
Note, collisional ionization is ruled out, as many of
the C IV profiles exhibit components with
$\sigma_{v}$ $<$ 5 {\kms} indicating $T$ $<$ 3{$\times$}10$^{4}$ K,
which is lower than the $T$ $>$ 4{$\times$}10$^{4}$ K 
threshold required 
to produce significant fractions of C$^{3+}$. In that respect
{\DLAs} differ from the ISM in which (a) C$^{3+}$ is collisionally
ionized in gas
with $T$$\sim$10$^{5}$K 
\citep{sembachs92}, 
and (b) the velocity components of the C IV and low-
ions are aligned. This is interpreted as evidence for
corotation of 
hot halo gas with the neutral disk
\citep*{savageed90}.
By contrast, the C$^{3+}$
ions in {\DLAs} are more likely produced 
in gas that is photoionized, has a temperature
$T$$\approx$10$^{4}$ K, and is  kinematically
distinct from the neutral gas. 

{\DLAS} also exhibit  Si IV $\lambda$$\lambda$1393.7, 1402.7 and
Al III $\lambda$$\lambda$ 1854.7, 1862.7
absorption. Because the Si IV and C IV velocity components
are closely aligned, the Si$^{+3}$ ions
must be produced by photoionization in the same
gas containing the C$^{3+}$ ions 
\citep{wolfep00a}.  
In the ISM the Al III lines arise in a warm
ionized medium (WIM), which is an extensive region of
warm ($T$$\approx$10$^{4}$ K) photoionized hydrogen
that pervades the disk of the Galaxy 
\citep{reynolds04}.
Therefore, it is surprising that in {\DLAs} 
the Al III velocity components are aligned
with the low ions and misaligned with the C IV and Si IV components
which arise in gas resembling a WIM.
This implies that the Al$^{2+}$ and low ions
both arise either in neutral gas or they
arise in photoionized gas, which is kinematically
distinct from the C$^{3+}$ bearing gas.
Photoionization equilibrium  calculations indicate
that in most {\DLAs} both Al$^{2+}$ and
the low ions arise in gas that is mainly neutral
\citep{vladiloetal00,prochaskaetal02b,wolfeetal04}.
The soft X-ray background
at $z \sim 3$ is sufficiently bright to 
produce
Al$^{2+}$ by photoionization and to leave hydrogen 
mainly neutral and carbon singly ionized 
(see 
\citealt{wolfeetal04}).
On the other hand 
in about 10 $\%$ of {\DLAs} the 
regions giving rise to low ion and Al III absorption
contain hydrogen that is significantly ionized. These 
cases can be recognized by the large  Fe$^{2+}$/Fe$^{+}$ ratios
and low Ar$^{0}$/Si$^{+}$ ratios 
\citep{prochaskaetal02b}.

These studies raise some interesting questions.
First, 
photoionization by X-ray background
radiation produces ions
X$^{j}$ for which the ionization potential of the
preceding
ionization state IP(X$^{j-1})$  $>$ 1 Ryd. 
In many {\DLAs} photoionization by locally generated
radiation with {\hnu} $<$ 1 Ryd 
produces
low ions for which 
IP(X$^{j-1})$ $<$ 1 Ryd (see $\S$~\ref{sect:sfr}).
The question is why doesn't local radiation with
$h{\nu}$ $\ge$ 1 Ryd 
photoionize sufficient neutral gas to
generate a co-rotating WIM like that in
the Galaxy? The answer may be related to the higher
neutral gas content of {\DLAs}, which could reduce
the escape fraction of 
ionizing radiation from H II regions in {\DLAs}.
Second, do {\DLAs} contain hot ($T$ $>$ 10$^{5}$ K)
collisionally ionized gas?  Hot gas is a byproduct of
feedback processes such as galactic winds or shock
heating by supernova remnants 
\citep{ferraras04}.
Because of the
evidence for type II supernovae 
(see $\S$~\ref{subsect:nucleosyn}) 
and the possible existence of winds
(see $\S$~\ref{sect:nums}), 
hot gas could be present in 
{\DLAs}. The most efficient technique for
finding hot gas is through the detection
of the 
O VI $\lambda\lambda$ 1031.9, 1037.6 doublet
\citep{sembachetal03}.
But no surveys
for 
O\,VI in {\DLAs} have been carried out.

\section{MOLECULAR GAS}

Molecular gas is ubiquitous throughout the Galaxy ISM.
In a survey for Lyman and Werner band absorption, the most
efficient tracers of {\H2} molecules in diffuse interstellar
gas,  
\citet{savageetal77} 
 detected {\H2} in 90 out of 
103 sightlines toward background Galactic stars. 
More specifically,
\citet{savageetal77}
 showed that the molecular fraction, 
$f$({\H2})
{$\equiv$}2$N$({\H2})/[2$N$({\H2})+{\nh}],  undergoes
a steep transition  from 
$f$({\H2}) $<$ 10$^{-4}$ at
{\nh} $<$ 4{$\times$}{\tenO} to
$f$({\H2}) $\ge$ 10$^{-2}$ at
{\nh}
$>$ 
4{$\times$}{\tenO}. 
Because of the similarity with the
{\nh}$\ge$2{$\times$}{\tenO} threshold, one might expect to
find $f$({\H2}) $>$ 10$^{-2}$ in a significant
fraction of {\DLAs}. Furthermore, 
if {\DLAs} are the neutral gas reservoirs for star formation
at high redshifts, 
and since stars form out of molecular clouds,
molecules should be present. 

However, the {\H2} content of {\DLAs} is much lower than in the Galaxy.
In their compilation of accurate
searches for {\H2},
\citet*{ledouxps03}
 report the detection of 
{\H2} in only five out of 23 cases qualifying as confirmed {\DLAs},
which brings to mind the {\H2} content of the LMC and
SMC. In particular
\citet{tumlinsonetal02}
found {\H2} in only 50$\%$ of the sightlines through the LMC. 
Moreover, the LMC resembles the {\DLAs} in that 
the typical upper limits are 
$f$({\H2}) $<$ 10$^{-5}$. 
In addition, the mean value of $f$({\H2}) for the
positive detections is about 10$^{-2}$ for the {\DLAs}, 
the LMC and the SMC, which is  about a factor of 10
lower than in the Galaxy.

Why is the {\H2} content in {\DLAs} so low?
The answer is partially related to low dust content.
In the ISM, {\H2} forms on the surfaces of dust grains
and is destroyed by photodissociation due to FUV radiation. In that
case $f$({\H2})=2$R$$n_{{\rm H}}$/$I$ 
\citep{jura74},
where $R$ 
is the formation rate constant and $I$ is
the photodissociation rate. Because
$R$
$\propto \ \kappa$$n_{H}^{2}$, $f$({\H2}) is predicted to decrease
with decreasing dust-to-gas ratio, and since $I$ $\propto$
{\jnu}, where {\jnu} is the mean intensity of FUV radiation,
$f$({\H2}) should  decrease with {\em increasing} radiation 
intensity. The low molecular fractions of the LMC and SMC
are plausibly attributed to 
low dust content 
and high radiation intensities. 
Similarly,
the low dust content of {\DLAs} helps to explain the
low values of $f$({\H2}). Indeed 
\citet*{ledouxbp02}
find a statistically significant positive correlation 
between $f$({\H2}) and {$\kappa$}. 

What is the value of
{\jnu} in {\DLAs}? 
\citet{levshakovetal02}
 inferred {\jnu}
for DLA0347$-$38
at
$z$ = 3.025 
by showing that FUV pumping was responsible for populating
several excited rotational levels in the ground electronic state.
By combining the excitation equations with the formation 
equations it is possible to deduce 
$I$, independent of the functional
form of $R$, and from that, {\jnu}. 
\citet{levshakovetal03}
found that {\jnu} was
comparable to the ambient interstellar radiation intensity
in the Galaxy, i.e., {\jnu}$\approx$10$^{-19}$ {\junit} for
this system.
Previous
authors reached similar conclusions for other {\DLAs} (see
\citealt*{blackcf87,petitjeansl00,geb97}).
 This is an important result, because 
intensities of this magnitude are 
significantly higher than the background
radiation intensity predicted
at {\hnu} $\approx$ 10 eV and $z$ $\approx$ 3 
\citep{haardtm03}.
Therefore, a local source of radiation is required to maintain
the right balance between {\H2} formation and destruction.



While the molecular content of the diffuse gas detected
in {\DLAs} is low, dense molecular clouds with high
dust content could be present. Such objects would
be missed owing to obscuration of the background
QSOs or due to a low covering factor. While future
surveys for radio-selected {\DLAs} may eventually
rule out such scenarios, they are consistent with
the current data. 
In fact dense molecular clouds may be required as the sites of the
star formation which has been  inferred for \DLAs.

\section{KINEMATICS}
\label{sect:kinematics}


As described in the previous sections, the introduction of echelle
spectrometers on 8 to 10 m-class telescopes has led to 
precise column density measurements of weak
transitions in the {\DLAs}.
It has also led to another significant advance in
damped {\lya} research, namely
the resolution of kinematic characteristics from 
the velocity profiles of the metal lines.
In $\S$~\ref{sect:mtlmeth} we showed that damped {\lya} profiles
exhibit multiple components that are qualitatively
similar to the component structure of similar
transitions observed in the Galaxy ISM \citep[e.g.][]{wolfeetal94}.
At high signal-to-noise ratios 
the line-profiles generally decompose into
5 to 30 velocity components, i.e., `clouds',
with column densities spanning roughly an order
of magnitude. The velocity components comprising 
the low-ion profiles (e.g.\ Si II $\lambda$ 1808.0)
are typically broadened by turbulent motions and have 
velocity dispersions of  $\sigma_{v}$ $\approx$ 4$-$7 {\kms}.  
These characteristics are also observed for the high-ion profiles
(e.g. C IV $\lambda$ 1548.1; 
\citealt{wolfep00a}), 
 although the
velocity dispersions are generally larger.

\citet{prochaskaw97}
presented the first modest sample of measurements
on the low-ion kinematic characteristics of the damped \lya\ systems.
Their results and subsequent surveys have shown that the {\DLAs}
exhibit velocity widths {\delv} ranging from 15 {\kms} to several 
hundred \kms\ with a median of $\approx$ 90 {\kms} 
(Figure~\ref{fig:lowkin}). 
\citet{prochaskaw97,prochaskaw98}
demonstrated that the observed {\delv}
distribution matched that predicted for rotating disks
with typical velocity speed $v_c \sim$ 200 {\kms} under the important
assumption that the gas disk is thick $(h  > 0.1 R_d)$,
where $h$ is the disk scale height and $R_{d}$ is the
radial exponential scale length.
The authors also stressed that the {\dla} line-profiles
tend to show the `edge-leading asymmetry' expected for rotating
disks.  The observations, therefore, suggested a population of disk
galaxies with rotation speeds corresponding to present-day galaxies.

The difficulty with this scenario, however, is that hierarchical
cosmology implies that galaxies in the young Universe have smaller
masses and lower rotation speeds
(on average) than the current population.  
Indeed, 
\citet{prochaskaw97} 
emphasized
that the circular velocity distribution of {\DLAs} predicted within the
CDM cosmogony 
\citep{kauffmann96} 
is incompatible with the
observations.  
This includes models (e.g.\ 
\citealt*{momw98})  
which allow for a cross-section weighted
distribution of spin parameters \citep{wolfep00a}.
As emphasized by \cite{jedamzikp98} and 
\cite{prochaskaw01}, there is a tension within CDM theory
between including enough low mass galactic halos to 
match the observed incidence of {\DLAs} without severely 
underpredicting the {\delv} distribution.
Granted the many successes of CDM theory, perhaps it 
will remain a true coincidence that the \dla\ kinematic characteristics
are best described by a population of galaxies similar to (albeit thicker
than)
 the present-day disk population. 

If we are to adopt the {$\Lambda$}CDM power spectrum
at $z \sim 3$,
then the velocity fields of a significant fraction of {\DLAs} must have 
contributions from non-rotational dynamics in order to 
explain the kinematic data.  
Numerical simulations within the CDM context
describe the {\DLAs} as multiple
`proto-galactic clumps' bound to a virialized dark matter halo
\citep*{haehneltsr98}.  The kinematics within this scenario
are due to the combination of infall,
random motions, and rotational dynamics.  In order for
this model to be consistent with the observations,
however, one requires that the {\dla} cross-section
$A$ is proportional to $v_c^\alpha$ 
with $\alpha > 2$ (\citealt*{haehneltsr98}; \citealt{malleretal01}).  
As discussed in $\S$~\ref{sect:nums}, early numerical work indicated 
$A \propto v_c^{1.1}$ which
implied a crisis between theory and observation \citep{prochaskaw01}.
More recent results, however, support $A \propto v_c^{2.5}$
\citep*{nagaminesh04a} and the protogalactic clump scenario remains
a viable option\footnote{We note that in the clump model the 
`edge-leading asymmetry' can be reproduced if 
3 or fewer
clumps are intercepted by the majority of {\dla} sightlines.}.  
We stress, however, that no cosmological simulation to date
has self-consistently matched the low-ion {\dla} kinematic observations.

\citet{luetal96} first remarked that the high-ion profiles (e.g. C\,IV
$\lambda$ 1548.1)
of the {\DLAs} have significantly different kinematic characteristics
from the low-ion gas.  
\citet{wolfep00a} examined a large sample of 
high-ion {\dla} profiles and stressed that while the component structure
is generally disjoint from the low-ion profiles,
the high-ions are roughly centered on the low-ion transitions.
Furthermore, there is a connection between the velocity fields 
of the two ionization states in that the high-ion gas nearly
always shows comparable or larger velocity width than the low-ion
gas (Figure~\ref{fig:kinciv}).  These trends place important constraints
on the nature of the {\DLAs}.
\citet{wolfep00b} considered a simple model where the low-ion gas
is confined to a disk enshrouded in a halo of high-ion gas
with kinematics described by the infall model of \citet{mom96}.
\citet{wolfep00b} demonstrated that this model could not simultaneously
match the low and high-ion kinematics, in particular the co-alignment
of the profile centers.  In contrast, 
\citet{malleretal03} demonstrated that a {\dla} model based on multiple
satellites bound to a single dark matter halo can satisfy the low-ion
and high-ion kinematics provided each satellite has a halo
of C IV producing  gas.  Their model only marginally reproduced the 
correlation between low-ion and high-ion gas and the authors suggested
that a kinematic correlation exists beyond their simple model.
Perhaps numerical simulations of the hot and cold phases will 
show that the hot gas co-rotates with the cold gas.

Because {\DLAs} are selected solely on the basis of H I  column
density, it is possible that other mechanisms contribute to their
velocity fields.  This could include winds induced by mergers or
star formation 
(\citealt*{nulsenbf98}; \citealt{schaye01}) 
 or even Hubble flow motions
in the ambient IGM.  At present
such scenarios have not been quantitatively developed or
tested against observation.  

One can gain further insight into the nature of {\DLAs} by 
synthesizing the 
results on kinematic characteristics with their other properties. 
\cite{wolfep98} first noted relationships between
the velocity width and the H I column density and metallicity
of the {\DLAs}.  
They found that the {\DLAs} with larger velocity width tend to
have higher metallicity.  
This trend matches one's physical intuition: If the 
velocity width is correlated with the galactic mass,
a correlation between \delv\ and [M/H] is natural provided more
massive galaxies have higher metallicity.
%
In contrast to the metallicity-{\delv} distribution, the
{\DLAs} show smaller {\nh} at large {\delv} values.
The observations run contrary to expectation for disk galaxies where
the gradient of rotation 
projected along the line of sight peaks toward the center.  The protogalactic
clump scenario also would predict a mild, positive correlation
\citep[e.g.][]{malleretal01} or no correlation depending on whether
H I surface density correlates with halo mass.  Therefore,
the {\dla} observations are
unexpected and difficult to interpet in terms of single or
multiple rotating disks. 
Additional studies, both observational and theoretical, on correlations
between the kinematics and other damped \lya\ properties 
would place further constraints on the processes of galaxy formation
in the young Universe.


It is important
to consider the implications
of the {\dla} kinematic characteristics with respect to the observed
relative abundance patterns (e.g.\ 
\citealt{prochaska03b}
).
As noted in $\S$~\ref{subsect:nucleosyn}, several authors have recently
argued that dust-corrected $\alpha$/Fe ratios of {\DLAs} are roughly
solar and
indicate star formation histories representative of dwarf or
irregular galaxies instead of massive systems 
(e.g. \citealt*{caluramv03}; \citealt{tolstoyetal03}).
We note, however, that the velocity widths
of the majority of {\DLAs} exceed 60~\kms\ and cannot be attributed
to the gravitational velocity field of a single dwarf galaxy.
Furthermore, there is no apparent correlation between the gas-phase
Si/Fe ratio and velocity width.  Therefore, a
contradiction exists between the observed 
velocity widths and the interpretation
of the {\dla} abundance patterns in terms of 
absorption by a single dwarf galaxy along the line of sight.
Regarding the CDM clump scenarios,
in which multiple dwarf galaxies are found
along the line of sight, one notes that the majority
of {\DLAs} are predicted to be embedded in dark matter halos which exceed 
the masses of present-day
dwarf galaxies 
(e.g. \citealt{malleretal01}; \citealt*{nagaminesh04a})  
and instead correspond to the
progenitors of galaxies like the Milky Way \citep{steinmetz03}.
It would be particularly valuable to perform a simulation which
traced the star formation history 
and resolved the galaxy kinematics within a cosmological context
to examine abundance pattern correlations with gas kinematics.

\section{GALAXY IDENTIFICATIONS}
\label{sect:ids}

In this section we describe the results of searches for
galaxies physically associated with {\DLAs}.


\subsection{Galaxies with $z$ $\ge$ 1.6}
\label{sect:highzids}

Because of the
presence of bright, $B$ 
$\sim$ 18.5, background QSOs, surveys for 
galaxies associated with high-$z$ {\DLAs} are
more challenging than surveys for randomly selected galaxies. 
To illustrate this
point consider a sightline passing through an $L_{*}$ galaxy\footnote{As 
defined in the $z=3$ Lyman Break Galaxy luminosity function 
of \citet{steideletal99}.}
at a reasonable impact parameter of 10 kpc. At $z$ = 3 the 
galaxy will have an AB magnitude
of 24.7 and impact parameter of
1.3 arcsec. Detection of the galaxy
against the QSO PSF with ground based telescopes would be exceedingly
difficult even under the best seeing conditions and has even 
proven difficult with broadband space imaging. Using the
NICMOS IR camera on the HST,  
\citet{colbertm02} surveyed 22 {\DLAs} and detected only one 
candidate counterpart down to $H_{AB}=23.5$, implying that most {\DLAs} 
are not drawn from the luminous end of the
Lyman Break Galaxy luminosity function.
\citet{warrenetal01} probed even deeper, to $H_{AB}=25$, and found 41 
candidate counterparts near 18 high-redshift \DLAs.  Broadband 
imaging to the necessary depth 
is thus limited by source confusion within reasonable impact 
parameters. 

The most widely used techniques are
searches for {\lya} emission lines at the
absorption redshift. The advantage of this method is that
the wavelength of 
{\lya} emission is located at the bottom of the damped
{\lya} absorption trough, which blocks the bright light of the
background QSO. As a result, background
night sky emission is the only source of external noise. 
Using slit spectroscopy, 
\citet*{foltzcw86}
failed to detect 
{\lya} emission
with a 
3-$\sigma$ upper limit of $F$ $\sim$ 10$^{-16}$
{\funit} for an unresolved object. 
While 
\citet*{hunsteadfp90} 
claimed detection of {\lya} emission from 
a compact source coinciding with
DLA0836$+$11
at $z$ = 2.466, this feature was not confirmed
in spectra acquired by 
\citet{wolfeetal92}
and 
\citet{lowenthaletal95}.
Nor was {\lya} emission detected in imaging surveys using narrow-band
interference filters or Fabry-Perot
interferometers. 
In this case
the QSO light is blocked because
the bandwidth of the filter 
is centered on the 
damped {\lya} line
but has a narrower FWHM. 
This technique is 
ideal for impact parameters large compared to
the seeing radius, since an emitter located outside a slit
could still be detected in the narrow band image.
\citet{smithetal89}, \citet*{deharvengbb90} and \citet{wolfeetal92}
carried out narrow-band surveys for {\lya} emission. 
\citet{lowenthaletal95}.
 carried out  Fabry-Perot surveys. 
No detections to
limiting fluxes of $F$$\approx$5{$\times$}10$^{-17}$ {\funit} 
for unresolved objects were reported. The
extended {\lya} emitter
associated with  DLA0836$+$11
at $z$ = 2.466
claimed by Wolfe {\etal} (1992) is more likely to be a galaxy
associated with a lower redshift Mg II absorption system
\citep{lowenthaletal95}.

The failure to detect {\lya} emission could result
from the destruction of resonantly trapped photons
by even a small amount of dust 
\citep{charlotf91}.
 For this reason several
groups attempted to detect {\DLAs}
in H$\alpha$ emission since H$\alpha$ photons 
are not resonantly trapped.
\citet{bunkeretal99}
 used IR detectors on a 4 m class
ground-based telescope to search for H$\alpha$ in five {\DLAs}, but 
none were detected (see also 
\citealt{mannuccietal98}).
Of course, the failure to detect H$\alpha$ emission might also be
due to small impact parameters, since an emitting region located within
the PSF of the QSO would not be detected from the ground.  
However, \citet{kulkarnietal00,kulkarnietal01}
used
NICMOS to search for H$\alpha$
emission from two {\DLAs} and none was found
(see Table~\ref{tab:emitters}). 

With the recent detections of {\lya} emission 
from at least 2 out of 18 {\DLAs} surveyed  
using 8-to 10-m class telescopes
(\citealt{molleretal02}; \citealt*{mollerff04}), 
it is evident that
the previous null detections of {\lya}  were largely due to the
lower sensitivity of 4-m class telescopes. The results,
summarized in Table~\ref{tab:emitters}, indicate that two of the three known
{\lya} emitters, DLA0458$-$02 and DLA0953$+$47A, would not
have been detected in the earlier surveys.
While the sample is too small to draw general conclusions,
the results are interesting for the following reasons.
First, {\DLAs} resemble randomly selected {\lya}
emitters by the similarity in {\lya} luminosity
and in the compact size of the emission regions.
Second, the small impact parameters for DLA0458$-$02
and DLA0953$+$47A suggest that the H I absorbing
layers are smaller than $\sim$ 5 kpc.
However, since the H I content of DLA0458$-$02 is known to extend
over linear scales exceeding 17 kpc 
\citep{briggsetal89}, 
 the sightline to the QSO must pass close to a 
compact star forming region, which
is embedded in a much larger layer of H I.
Third,
no continuum emission has been detected from 
the same two  {\DLAs}. The  $B$ $>$ 27 limit on
DLA0953$+$47A  
places
this {\DLA}  near the faint end of the known luminosity
function of Lyman Break galaxies. 

By contrast, DLA2206$-$19A is a luminous Lyman Break galaxy.
This emitter was first located in IR images
obtained with NICMOS 
\citep{warrenetal01}.
 The 
STIS image  \citep{molleretal02}  shows rest-frame
FUV stellar emission extending 1 arcsec between
a bright knot  and the QSO sightline. {\lya} emission
at the redshift of the {\DLA} was detected from
the knot with spectra obtained with the VLT 
\citep{molleretal02}.
The magnitude integrated over the object is $V$= 23 
(M$\o$ller 2004, priv. comm.), 
which places this {\DLA}
at the bright end of the Lyman Break luminosity 
function. 

As a result, there is
little overlap between the luminosity
functions of {\DLAs} and the $R$ $<$ 25.5
spectroscopic sample of Lyman Break galaxies
(see 
\citealt{molleretal02}; \citealt{schaye01})
Efforts to detect the clustering of {\DLAs} with neighboring 
Lyman Break galaxies
have so far only yielded upper limits \citep{gawiseretal01,adelbergeretal03}, 
providing further evidence that the {\DLAs} are nearly disjoint 
from the $R<25.5$ spectroscopic sample of 
Lyman Break Galaxies, with DLA2206$-$19A 
a clear exception. 
 Based on the UV continua implied by the 
{\lya} luminosities of the other 2 {\DLAs} detected in emission, there is 
growing evidence that at least 
some {\DLAs} overlap with the dimmer ``photometric'' 
sample of Lyman Break Galaxies at $R<27$ seen in the Hubble Deep Fields.  

\subsection{Galaxies with $z$ $<$ 1.6}
\label{sect:lowzids}

While the nature of the galaxies associated with {\DLAs}
at $z$ $>$ 1.6 is still unclear, at lower redshifts
there should be a close resemblance to objects drawn from the
population of normal galaxies 
if {\DLAs} trace the star formation history
of normal galaxies. However, the low metallicities inferred
for most low-$z$ {\DLAs} 
\citep{pettinietal99,kulkarnietal05}
has cast doubt on this idea 
and has
led to the suggestion that low-$z$ {\DLAs} are metal-poor
objects such as dwarf galaxies 
\citep*{caluramv03}
or
low surface-brightness galaxies 
\citep*{jimenezbm99}.
As a result, identification of galaxies associated with
{\DLAs} at $z$ $<$ 1 is of vital importance.

There are 23 {\DLAs} known at $z < 1.6$ (as of October 2004).
Thirteen of these have been identified using spectroscopic
or photometric redshift techniques 
\citep*{chenl03,chenkr05}
 and 
possible host galaxies for an 
additional eight systems have been found in imaging
surveys 
\citep{lebrunetal97,raoetal03}.
\citet{chenl03}
 analyzed an unbiased
subset of nine  galaxies with redshifts and concluded
that the resultant luminosity distribution  was dominated by
luminous galaxies with $L/L_{*}$$>$ 0.1; i.e., luminous
galaxies dominate the neutral gas cross-section at $z$ $<$ 1.
More specifically, they used a maximum-likelihood technique
to determine the dependence of H I cross section on luminosity
and showed that a cross-section weighted Schechter function
with typical parameters for normal galaxies provided a good fit
to the data.

On the other hand, 
\citet{raoetal03}
 analyzed a heterogeneous
sample of fourteen 
{\DLA} host galaxy candidates, including objects without confirmed
redshifts and concluded that
the neutral gas cross-section at $z$ $<$ 1 was dominated by
dwarf galaxies. Since
\citet{raoetal03}
 did not
quantify their result, this conclusion is difficult to evaluate. 
However, comparison with the sample of \citet{chenl03} shows a larger
fraction of galaxies with lower values of $L/L_{*}$ in the
\citet{raoetal03}
 sample. In some cases these are galaxies 
without confirmed redshifts at small
angular separations from the damped {\lya} sightline. In other
instances, where  two or more galaxies are found to be associated
with the {\DLA},
\citet{raoetal03}
 chose the lowest luminosity
galaxy for the analysis because it had the smallest impact
parameter. However,
\citet{chenl03}
argue that in
these cases, an entire group of galaxies is responsible for
the absorption profile, so the object with lowest impact parameter
may not be the appropriate choice. 
Given these uncertainties and the  small sizes of the samples,
we conclude that the current data 
are consistent with the galaxies associated with low
redshift {\DLAs} being drawn from the population of
normal galaxies.

\section{STAR FORMATION IN DAMPED {\lya} SYSTEMS}
\label{sect:sfr}

Several lines of evidence imply that {\DLAs}
experience ongoing star formation. At $z$ $<$ 1.6 the evidence
is unambiguous, since
more than half  the sample 
is associated
with galaxies of stars. At $z>1.6$ 
the evidence is 
indirect
because
only one {\DLA} 
is identified with a resolved object
of galactic dimensions
that emits starlight.  While starlight likely
ionizes the gas that gives rise
to {\lya} emission 
in two other high-$z$ {\DLAs},
the starlight itself has not been directly
detected.
The expectation is that most of the metals seen in 
{\DLAs} were generated through star formation in their host systems, 
but the instantaneous star formation rates 
characterizing  this important population of objects remained unknown
until recently.  

\subsection{Direct Emission Measurements of 
Star Formation Rates in Damped {\lya} Systems with $z$ $>$ 1.6}  
\label{sect:sfrdirect}


Table~\ref{tab:emitters} summarizes results including 
SFRs for the three  positive detections and 
for  objects with upper limits on SFR at $z>1.6$.  
Observed lower limits on 
$L$(H$\alpha$) were coverted to upper limits on SFR using 
the \citet{kennicutt98}
calibration, 
SFR(M$_{\odot}$yr$^{-1}$)=7.9${\times}10^{-42}$$L({\rm H}{\alpha})({\rm ergs \ s^{-1}})$.
Detections of {\lya} emission\footnote{Due to its extreme 
sensitivity to dust extinction, {\lya} emission 
provides a lower limit to 
the star formation rate as long as the AGN contribution to the emission 
line is negligible.}   
were combined with the expression for $L$(H{$\alpha$})
and case B radiative recombination
to find lower limits of 
SFR(M$_{\odot}$yr$^{-1}$)=1.1${\times}10^{-42}$$L({\rm Ly}{\alpha})({\rm ergs \ s^{-1}})$.
The total magnitude integrated over DLA2206$-$19A of $V$= 23  
and the very sensitive $B$ $>$ 27 limit on
DLA0953$+$47A 
(A. Bunker 2004, priv. comm.)
were combined with 
the rest-frame UV 
\citet{kennicutt98} 
calibration,
SFR(M$_{\odot}$yr$^{-1}$)=1.4${\times}10^{-28}$$L_{\nu}({\rm ergs \ s^{-1} \ Hz^{-1}})$,
to set a lower limit on the SFR for DLA2206$-$19A and a rough upper limit 
for DLA0953$+$47A assuming no dust extinction for the latter object.  
An estimated dust extinction correction for DLA2206$-$19A 
\citep{wolfeetal04},
suggests an upper limit on SFR of 50 {\smpy},  
which 
is comparable to the
SFRs of the more luminous Lyman Break galaxies 
\citep{shapleyetal03}.

%

\subsection{Star Formation Rates from the C II$^{*}$ Technique}
\label{sect:cii*}

A method recently developed by \citet*{wolfepg03} makes it possible 
to infer the star formation rate per unit
area for individual 
{\DLAs}.  The basic method is to use measurements
of the C II$^*$ 1335.7 
column density 
 to measure 
the [C II] 158 {\micron} cooling 
rate in the neutral gas producing the {\dla} absorption.  
This is possible because the {\ciis} $\lambda$ 1335.7
transition arises from the excited $^{2}P_{3/2}$ state 
in C$^{+}$, and spontaneous photon
decay of 
the $^{2}P_{3/2}$ state to 
the $^{2}P_{1/2}$ state results in
[C II] 158 {\micron} emission,
which is the principal coolant of neutral
gas in the Galactic ISM 
\citep{wrightetal91}.
Under the
presumed condition of thermal balance,
the cooling rate equals the heating rate and it is possible to
calculate the star formation 
rate per unit area that generates the implied heating rate.

By analogy with the Galactic ISM,
\citet*{wolfepg03}
adopt the grain photoelectric
effect as 
the principal
heating mechanism for {\DLAs}. In that case  FUV radiation
({\hnu}$\approx$ 6 to 13.6 eV)
ejects
photoelectrons from grain surfaces, which heat ambient
electrons through Coulomb interactions 
\citep{bakest94,weingartnerd03}.
The 
heating rate from the grain photoelectric effect is proportional 
to the FUV radiation intensity {\jnu}, which consists of a 
contribution from the FUV background radiation plus a local contribution 
from 
hot stars located inside the galaxy
which 
is proportional to
the instantaneous star formation rate per unit area, {\ps}.

To infer \ps}  and other properties from observations,
\citet*{wolfepg03}
deduce the 
[C II] 158 {\micron} spontaneous emission rate 
per H atom from the observational
quantity:
\begin{equation}
\ell_c = \frac{N(\mathrm{C \; II}^*)h \nu_{ul} A_{ul}}{N(\mathrm{H \; I})}
    \mathrm{ergs \; s^{-1} H^{-1}} , \; \; \; 
\label{eq:lc}
\end{equation}
where $A_{ul}$ is the Einstein coefficient and $h{\nu_{ul}}$ is 
the energy of the 158 $\mu$m transition.  These are known quantities, 
and $N(\mathrm {C \; II}^*)$ and $N(\mathrm{ H \; I})$ 
are measured from the 
absorption line spectra.  
The total heating rate 
includes 
inputs due to  the grain photoelectric 
effect, X-ray photoionization, cosmic ray ionization,
C I photoionization, and collisional heating.  
The total cooling rate includes  
cooling due to {\lya}, grain radiative
recombination, and emission by fine-structure states
of O$^{0}$, Si$^{+}$, Fe$^{+}$, etc. in addition to
the [C II] 158 {\micron} fine-structure emission. 
Radiative excitation of the C$^{+}$ fine-structure
states by CMB radiation 
is included because it 
can be significant at high redshift. 
\citet{wolfeetal04} 
showed that the heating rates
predicted for the 
\citet{haardtm03}
backgrounds
were significantly lower than the 158 {\micron} cooling rates
implied for {\DLAs}
with detected {\ciis} $\lambda$ 1335.7 absorption, requiring 
active star formation to explain the observed cooling rates.  
On the other
hand, 
\citet{wolfeetal04}
also showed that
heating by background radiation alone cannot be
ruled out for 
systems in which {\ciis} $\lambda$ 1335.7 absorption
was not detected. A summary of the {\lclos} data for
{\DLAs} (and the Galaxy) is given in 
Figure~\ref{fig_lcvsNHI}. 
\citet*{wolfepg03} argued that the heating rate in the Galaxy is
much larger than in {\DLAs} because the dust content
in the Galaxy is at least a factor of 30 
higher than in {\DLAs}, whereas {\ps} in the Galaxy 
is only a factor of 2 to 3 lower than in {\DLAs}
(see below).

In the case of systems with detected {\ciis} $\lambda$ 1335.7 absorption,
one compares 158 {\micron} emission rates computed for a 
range of 
star formation rates 
per unit area 
with the empirical quantity, {\lclos}. This results in 
in 
two solutions for {\ps}
corresponding to thermally stable
states of a two-phase medium in which
warm neutral medium (WNM) gas is in pressure equilibrium
with cold neutral medium (CNM) gas: the WNM solution corresponds
to gas
with
$T$$\sim$8000K and $n$$\sim$0.02 cm$^{-3}$, and the CNM solution corresponds
to gas with
$T$$\sim$100K and $n$$\sim$10cm$^{-3}$.
In every case 
the WNM solution for {\ps} 
is a factor of ten or more higher than the 
CNM solution:
since
C II emission in the WNM is a small fraction of the total
cooling rate, the total heating rate implied for an observed
value of {\lclos} must be higher in the WNM than in the
CNM. Owing to the low dust optical depths of 
{\DLAs} the resulting {\ps} are system-wide average
star formation rates rather than local values as
is the case in the dusty Galactic ISM.

This method was used by 
\citet{wolfeetal04}
to analyze a sample of 
45 {\DLAs}, 23 
with 
measured C II* column densities and
22 for which {\ciis} was not detected.
In the
redshift interval $z$ =[1.6,4.5] the average SFR
per unit area 
for positive detections 
 is {\psav} = 
11.3
{$\times$}10$^{-3}$ {\smpykpc}
for the CNM solution 
and
0.21
{\smpykpc} for the WNM solution; 
by comparison in the Galaxy {\ps}
$\approx$ 4{$\times$}10$^{-3}$ {\smpykpc}. However, the WNM solution
is unlikely to be correct, because 
the bolometric background intensity produced 
exceeds the 
observational limits 
\citep*[see][]{wolfegp03} .
This conclusion is consistent with recent evidence from individual 
systems; 
\citet*{howkwp05}
showed that the low upper limit on
the optical-depth ratio of Si II$^{*}$
$\lambda$ 1264.7 to {\ciis} $\lambda$ 1335.7
in a high-redshift {\DLA} results in an upper limit
of 800 K for the 
temperature in the gas producing {\ciis}, which
implies that {\ciis} absorption arises in CNM gas in this system.
Furthermore, a pure WNM solution would significantly 
overpredict the observed rest-frame-UV luminosity 
from DLA2206$-$19A based on its measured 
{\lclos}, implying that {\ciis} absorption in this
{\DLA} arises in a CNM.  
 It is significant that
the predicted value for CNM, 
{\jnu}$^{{\rm CII}^{*}}$ = 1.7$^{+2.7}_{-1.0}${$\times$}10$^{-18}$
{\junit} is the largest mean intensity inferred from the 
entire {\ciis} sample. This may help to explain why 
DLA2206$-$19A
is one of the rare {\DLAs} detected in emission.

On the other hand the gas detected in absorption is more likely
to be a WNM for {\DLAs} with upper limits on {\ciis} absorption.
In this case the gas could be a pure single-phase WNM
heated by background radiation alone. These
{\DLAs} would then be objects without significant star formation
at the absorption epochs. Or the gas could be the WNM branch
of a two-phase medium in which the SFR per unit area 
is similar to that found for the CNM solutions
\citep{wolfeetal04}. 
While the absence of 21 cm absorption at $z$ $>$ 3
\citep{kanekarc03}
and the large C II/C I ratios
\citep{liszt02}
are consistent with the WNM hypothesis, both phenomena
are also naturally explained within the context of the
two-phase hypothesis. 
Clearly, it is
important to determine
which of these explanations is the correct one.

\subsection{Implications} 
\label{sect:sfrimplications}



The {\ciis} technique can also be used to obtain quantities
with cosmological significance. Specifically, 
the SFR per unit comoving volume is
given by
$\dot{\rho_*}=\langle \dot{\psi}_*(z)\rangle (A_*/A_p)
(d{\cal N}/dX) (H_0 / c)$, 
where $A_{*}$ is the intrinsic cross-sectional area occupied
by stars emitting FUV radiation and
$A_{p}$ is the projection of the intrinsic H I area,
$A$, on the 
sky.\footnote{Note that $A_{p}$ equals $A(N,X)$ (see eq. 
\ref{eq:dN(n,z)}
) averaged over the
column-density interval {\nh} = [$N_{min},N_{max}$]}
In the uniform disk model, neutral gas and stars are uniformly
distributed across {$A$}, 
yielding 
the values for {\rhodotz} given 
in Table \ref{tab:global}, which also summarizes the systematic 
uncertainties  
discussed at length
 in 
\citet*{wolfegp03}.

The total mass of stars produced in {\DLAs} 
as 
a function of redshift 
can be computed by integrating over the cosmic star 
formation 
history of {\DLAs}.  
The mass density 
of stars predicted by today, including the contribution from ``normal'' 
galaxies at $z<1.6$,  is consistent with the total current 
density of stars in spiral and elliptical galaxies, but this 
does not reveal the precise population of stars produced by 
high-redshift {\DLAs}. The same star formation history will also
deplete the neutral gas reservoir at high redshift in about $t_{*}\approx 2$ 
Gyr; i.e., {\omgas} would vanish by $z \approx 2$ if star formation
starts at $z \approx  5$. Instead {\omgas} drops by a factor of two during
this time, which argues for replenishment of neutral gas at
an accretion rate {$\dot{\rho_{a}}$$(z)$}=0.5{\rhodotz}.

\citet*{wolfegp03}
 also calculated 
the mass density of metals produced in {\DLAs} by 
using the conversion of star formation rate to metal formation 
rate suggested by \citet{madauetal96} and \citet{pettini99},  
$\dot{\rho}_{metals} = (1/42) \dot{\rho}_* $.  
Extrapolating to the present day, including the contribution 
from star formation in  
normal galaxies at $z<1.6$, yields a mass of metals a few times 
larger than that found in spiral bulges today, which seems 
feasible.  
Integrating under the 
DLA cosmic star formation history 
at $z>2.5$ 
predicts a factor of 30 higher cosmic mean metallicity due to 
metal enrichment of neutral 
gas than is observed in {\DLAs} at $z=2.5$ 
\citep{prochaskaetal03b}.  
This sort of ``missing metals'' problem was
first identified by 
\citet{pettini99} 
for Lyman Break Galaxy star formation rates compared to {\DLA} metallicities 
but that problem could be solved by assuming that \DLAs\
are not the descendants of the star-forming Lyman Break Galaxies.  
The {\DLA} ``missing metals'' problem appears to be fundamental 
and may illuminate a fundamental flaw in our understanding of 
the relationship between metal enrichment and star formation at 
high redshift.    
The problem is a significant challenge not only to the 
C II$^*$ method of measuring {\dla} star formation rates, but 
to most hierarchical models, which 
also produce too many
metals (see 
\citealt*{nagaminesh04b,somervillepf01})\footnote{The model of 
\citet*{peifh99} 
avoids this problem by
invoking significant obscuration corrections and low yields. 
}. 
Given the 
other successes of the method it seems likely that the problem 
has a physical solution.

The simplest solution would be to hypothesize that the {\DLAs} forming stars 
at $z>2.5$ are no longer {\DLAs} at $z=2.5$ but have used up most of their 
neutral gas and now have the expected metallicities; the Lyman Break 
galaxies are an obvious candidate for their descendants.  
This is inconsistent, however, with 
timescale of $\approx 2$ Gyr to use up most of the neutral gas
implied by the observed star formation rates, which is too long.  
Ejecting the metals into the IGM 
does not solve the problem, 
as the IGM metallicity predicted at $z=2.5$ would be 
[M/H]$=-1.5$ i.e. about a factor of 30 
higher than observed in the Lyman $\alpha$ forest, 
unless 
the  ejected
gas is so hot that most metals are in an unobservable ionization
state, with the possible exception of oxygen, which
might be detected in O VI $\lambda$$\lambda$1031.9, 1037.6.  
An alternative possibility is that the
metals are sequestered, with the most attractive solution being 
a model in which {\DLAs} represent neutral gas on the outskirts of 
actively star-forming ``bulge'' regions with the majority of the 
produced metals remaining in these regions rather than polluting 
the {\dla} gas or general IGM \citep*{wolfegp03}.  
This hypothesis necessitates 
the conclusion that metal-rich bulge regions are underrepresented 
(or completely unrepresented)  
in the HI-weighted cosmic mean metallicity measured from DLAs.
That would not be surprising given that the 
bulges 
may well have used up 
most of their neutral gas and/or have sufficient dust content to 
dim background QSOs out of optically selected samples.

\section{CHEMICAL EVOLUTION MODELS for DAMPED Ly{$\alpha$} SYSTEMS}
\label{sect:chemev}


In the ``global'' approximation,
one averages 
 all quantities
over large comoving volumes and then solves
the chemical
evolution equations to deduce
metal production rates from the comoving 
SFR, {\rhodotz}, which is computed by tracking changes in 
the neutral gas density, {\rhog} 
\citep{lanzettawt95, peif95,malaneyc96, 
edmundsp97}. 
However, the factor-of-two decrease between $z$ $\approx$ 4 and
$z$ $\approx$ 2 suggests
that
{\DLAs} are replenished by a net inflow of neutral gas
(see $\S$~\ref{sect:sfrimplications}).


\citet{peif95}  considered models with inflow and outflow 
and searched for self-consistent evolution of the neutral gas, metallicity, 
and dust. 
These authors calculated the effects of obscuration 
on quantities deduced directly from the data, such as {\nh}, 
and 
found that significant obscuration was necessary to explain the observations.  
\citet{peif95}  
fitted the available data with analytic functions for {\rhog}
that increased with
$z$ at $z$ $<$ 2 {\citep*{lanzettawt95}} and assumed that
the net accretion rate, {\rhodotacc}, 
was proportional to {\rhodotz}. When more accurate
measurements of {\rhodotz} {\citep{steideletal99}}
and  the cosmic background radiation intensity {\citep{hauserd01}}
became available, \citet*{peifh99} also included the production of
background radiation by stars in {\DLAs}
(see \citealt*{fallcp96}). With these additional constraints 
\citet*{peifh99} 
worked directly from the
measurements of {\rhog} and 
eliminated the assumption that {\rhodotacc}
was proportional to {\rhodotz}.
The newer models reproduced the 
more accurate metallicity
measurements not available earlier,
and were consistent with measurements of the background
radiation intensities and {\rhodotz}.
However, 
as we discuss in $\S$ \ref{sect:dust}, 
these models appear to 
cause more obscuration than the current 
observational data imply. 
Furthermore, the values of {\rhodotz} 
at $z$ $<$2
were inferred from changes in {\rhog} that now 
appear to be spurious (Rao, Turnshek, \& Nestor 2004; priv. comm.). 

In the ``local'' approximation,
one computes the chemical evolution of isolated galaxies
outside  a cosmological setting.
A star-formation history is imposed from the outset
and one solves for the chemical response of stars and gas.
Adopting the slow star formation
history for spiral galaxies suggested by Mateucci, Molaro, \& Vladilo (1997),
\citet*{lindnerff99} reproduced the slow evolution
of [Zn/H] with $z$ observed in {\DLAs}. However, these authors
adopted a  ``closed box'' model and neglected
spatial gradients in all physical parameters. 
\citet*{caluramv03} removed these restrictions 
and also 
reproduced the slow increase of [Zn/H] with decreasing redshift using 
star formation histories
predicted for large galactic
radii in spirals or for episodic 
bursts in dwarf irregulars. Furthermore,
they used the same models to reproduce the
[Si/Fe] versus [Fe/H] relation after correcting for depletion.
\cite{dessaugeszavadskyetal04} used these models
to explain the chemical evolution of
three {\DLAs} for which abundances of a large number of elements
had been obtained. 
Interestingly,
the predicted SFRs per unit area agree with those
inferred from the {\ciis} technique. 
Whereas 
\citet*{wolfegp03} found that integrating
such SFRs between $z$ = 5 and 3 resulted
in the overproduction of metals, \cite{dessaugeszavadskyetal04}
found that the cumulative metals produced did not exceed
those observed owing to the short time-scales
for metal production required to explain
relative abundance
ratios such as [Si/Fe]. 
However, in some cases the short time scales 
conflict with the conservative
lower limit of 0.25 Gyr on age set by the
measurement of [N/$\alpha$]
near the $-$0.7 plateau (see $\S$~\ref{subsect:nucleosyn}). 
Other potential problems 
with these 
models 
stem from 
the depletion corrections
applied to the [Si/Fe] ratio, which may be 
too large (see $\S$~\ref{sect:dust}).



The most promising approach 
to chemical evolution is the direct one, which uses
cosmological hydrodynamical simulations (see 
\citealt*{somervillepf01} and \citealt{mathlinetal01} for
semi-analytic and analytic variants of this method).
The simulations unite the ``local'' and ``global'' 
approximations with a self-consistent evolution
of stars, gas, metals, and dust within a
{\lcdm} cosmology. 
While the microphysics behind star 
formation and metal production cannot be included in these simulations, 
recipes calibrated to local observations are used to track stars  
and metals along with dark matter particles governed by gravity 
and gas particles governed by gravity and hydrodynamics. As a result,
processes such as 
accretion of neutral gas from the IGM are described physically, and
star formation is treated self-consistently rather than being imposed
{\em ad hoc}. Moreover, merging between dark-matter halos is
included for the first time.

\citet{ceno99} were the first to describe the chemical evolution 
of {\DLAs} with numerical simulations. Using a low-resolution Eulerian scheme,
these authors
were unable to resolve the dark-matter halos giving rise 
to {\dla} absorption.
Nevertheless, they pointed out that metallicity
is a more sensitive function of overdensity, $\delta$, than of age:
metal-poor objects such as the {\lya}-forest clouds 
formed in low density environments ($\delta$ $\approx$ 1), 
while more metal-rich objects such
as {\DLAs} and Lyman Break Galaxies formed in regions of higher
overdensity ($\delta$ $>$ 10). Using a more accurate version of this
numerical code, 
\citet{cenetal03} 
predicted the cosmic
metallicity at $z \sim 3$ to be 
between 0.3 and 0.5 dex higher than the observed value. 
They solved this  ``missing metals
problem'' (see  $\S$~\ref{sect:sfrimplications})
by using obscuration
corrections that may be larger than
allowed by the results of \citet{murphyl04}.
They also predicted that, independent of metallicity,
the ages of typical {\DLAs} in the redshift
interval $z$=[2,4] would be 0.8$-$2 Gyr, which are consistent
with the presence of the upper [N/$\alpha$] plateau. 
Another
prediction of interest is that the median stellar
mass $M_{*}$$\sim$10$^{9}$ {\msolar}, which is a factor of 10 lower
than that of Lyman Break galaxies, indicating they are different
populations. 

The present state-of-the-art in cosmological 
hydrodynamic simulations of {\DLAs} is represented by the recent 
results of \citet*{nagaminesh04b} who used the SPH  
code described in $\S$~\ref{sect:nums} (see also \citealt{coraetal03}). 
These authors found SFRs per unit area that agree
with
the predictions of the {\ciis} technique for {\DLAs} \citep*{wolfepg03}.
They also predicted an overproduction of metals
by $z$$\approx$2.5, but in this case by a factor of 10 compared
to the observed metal abundances. The difference with \citet{cenetal03}
is likely related to the lower spatial resolution of the latter
simulation (about 30$h^{-1}$ kpc comoving), which causes 
the high-metallicities of compact regions 
to be diluted by the low metallicities of diffuse
regions.
One of the interesting predictions of the   
\citet*{nagaminesh04b} simulations is that all regions in which
{\nh} $\ge$ {\NL} exhibit star formation. Confirmation of this
prediction would favor the uniform disk model over the bulge
model of star formation discussed by \citet*{wolfegp03}. As a result,
it is important to decide whether this finding is an artifact of
the star formation algorithm employed by \citet*{nagaminesh04b},
especially since there are regions in nearby galaxies in which
{\nh} $\ge$ {\NL}, but only low  star formation rates ($\sim$10$^{-5}$
{\smpykpc}) are observed \cite{fergusonetal98}.


\section{ARE DAMPED {\lya} SAMPLES BIASED BY DUST?}
\label{sect:dust}

%

Surveys for {\DLAs} 
have the greatest impact if they
represent a fair sample of the neutral gas in the Universe, allowing a 
clear probe of the evolution with redshift of the neutral hydrogen 
content and the metallicity of neutral gas.
However, it has long been a major concern that the 
sample of {\DLAs} suffers from ``dust bias'' i.e. the absence from a 
magnitude-limited QSO sample of those QSOs which suffer obscuration 
from dusty foreground {\DLAs},
leading to underrepresentation of dusty {\DLAs} in the 
overall sample.  
The easiest way to probe the existence and abundance of dust in {\DLAs} would 
be to find the 2175{\AA} bump feature superimposed in absorption on 
background QSO spectra.  While at least one strong example has been 
found (Junkkarinen {\etal} 2004),
this does not appear to be 
the rule \citep*{peifb91}.
Without such sharp features to look for and given the 
wide range of intrinsic QSO spectral slopes, reddening from dust 
in {\DLAs} must be searched for statistically by checking if the 
sample of QSOs with foreground {\DLAs} is redder on average than 
a ``control sample'' of QSOs without foreground {\DLAs}.

\subsection{Observational estimates of reddening}

\citet{ostrikerh84} pointed out that optically-selected QSO 
samples are biased towards those QSOs with little foreground 
dust extinction.  
\citet{fallp89} 
showed that 
dust in {\DLAs} did not appear to cause the 
famous drop in the QSO number abundance at $z>3$.  
\citet*{peifb91} 
detected reddening from {\DLAs} 
at the $4 \sigma$ confidence level 
and inferred 
dust-to-gas ratios 
between 1/20 and 1/5 that 
of the Galaxy, enough to explain the lack of observed Lyman $\alpha$ 
emission from {\DLAs}.  This led to the prediction 
that 10\%-70\% of QSOs are missing from optically-selected 
samples, leading to an order of magnitude  uncertainty in $\Omega_{g}$, 
$<Z>$, and other quantities estimated from {\DLAs} \citep{fallp93}.  
However, the dust-to-gas ratios estimated from high-resolution 
echelle spectroscopy of QSOs with foreground {\DLAs} are lower than 
the dust-to-gas ratios predicted by 
\citet{fallp93}, 
reducing the 
uncertainty in quantities 
such as $\Omega_{g}$ 
to factors of 2-3.
\citet{pettinietal97a} combined a metallicity 
of 1/15 solar with a dust-to-metals ratio of 1/2 that in the 
Milky Way to find a typical {\DLA} dust-to-gas ratio of 1/30 Galactic.  
Using an SMC reddening curve,
they predicted a dust extinction of only 0.1 magnitudes at 1500{\AA} in 
the spectrum of background QSOs due to {\DLA} dust.  
If a nucleosynthetic floor exists in {\DLAs} at [Si/Fe]$\simeq 0.3$
then the dust-to-gas ratios are even lower than this, closer to 
1/200 Galactic in most systems.   
Indeed, the detection of reddening due to  {\DLAs} 
by \citet*{peifb91}
conflicts with the recent finding by 
\citet{murphyl04} that $E(B-V)<0.01$ magnitudes using 81 {\DLAs} found 
in a homogeneous set of SDSS Data Release 2 QSOs.  The resolution 
of the conflict is not clear at present.

\subsection{Surveys of radio-selected QSOs}


An insidious (but not physically motivated) possibility would be 
the existence of gray dust associated with the {\DLAs} which could 
cause obscuration without the telltale effect of reddening.  Even this effect
could be overcome by using a radio-selected sample of QSOs.  
The reason this has not typically been done is two-fold:  (1) 
The ability to select QSOs 
within a preferred redshift range 
makes 
 optical color selection 
more efficient.
(2)  Conducting 
optical spectroscopic follow-up on a radio-selected sample of QSOs 
is far more time-consuming precisely because they do not have a strict 
optical magnitude limit; half of the total exposure time can be required 
by the dimmest one or two objects.  Obviously, dropping those from the 
survey would defeat the entire purpose of radio selection.  
One radio-selected sample has been published:  
the CORALS survey of \citet{ellisonetal01b} 
found  
19 intervening {\DLAs} towards 66 $z_{em}\geq 2.2$ radio-selected 
QSOs from the Parkes quarter-Jansky sample \citep{shaveretal96}, 
yielding a marginal  
increase in {\dndx} and $\Omega_{g}$ at $<z>=2.37$ versus 
optically-selected QSO samples.  
These results imply 
that at most half of {\DLAs} are missing from 
optically-selected samples.  For $\Omega_{g}$, the 
radio sample yields 1.4$\times 10^{-3}$ 
as opposed to the value of 
6.7$\times 10^{-4}$ found for optically selected samples 
at this redshift \citep{prochaskah04}, 
but this is only 
a 1.5 $\sigma$ difference
given the small sample size.

\subsection{Empirical Estimates of Damped {\lya} System Obscuration} 

A third way to estimate the effects of dust obscuration by {\DLAs} is 
to infer this from the observed chemical abundances.   
%
Taking the observed H\,I column densities and the
dust-to-gas ratios implied by the depletion
patterns of the {\DLAs} (see Eq.~\ref{eq:kappa}), it is possible to estimate the extinction
in the rest-frame UV of the QSO 
for an assumed 
extinction
curve.
Given the lack of the observed 2175{\AA} bump feature, it appears 
more reasonable to assume an SMC \citep{prevotetal84} 
rather than Galactic \citep*{cardellicm89} dust extinction law.
\citet{prochaskaw02} used this technique (see their Figure~24; see
\citealt{prochaska04} 
for an update)
to correct the observed QSO magnitudes by this inferred extinction 
and then to compare the implied true magnitude with the magnitude 
limit of the survey used to search for {\DLAs} (which is typically 
shallower than the limit of the survey used to discover the QSOs).  
These quantities were then compared to a bootstrap prediction of 
how many QSOs are expected to be missing from observed samples 
due to extinction by foreground {\DLAs}.  The typical range of 
extinction corrections runs from 
0 to 0.3 magnitudes,  
even though half  
of the QSOs are so much brighter than the survey limit that 
they could have been seen with up to 1 magnitude of extinction.  
This shows that {\DLAs} which cause between 0.3 magnitudes 
and 1 magnitude of extinction are rare and 
predicts that at most 10\% of QSOs are missing from the 
samples probed for {\DLAs} due to a {\dla} dust bias.

\section{ARE DAMPED {\lya} SAMPLES BIASED BY GRAVITATIONAL LENSING ?}

On the other hand Prochaska, Herbert-Fort, \& Wolfe (2005) 
find that biasing due to 
gravitational lensing could be important. They compared the
full SDSS sample of {\DLAs} with subsamples comprising the 
brightest 33 $\%$ of background quasars and the faintest 33 $\%$
of background quasars. While the incidence of {\DLAs}, {\dNdX}
was found to be insensitive to quasar magnitude, the mass density,
{\omgas}, was found to vary significantly. Specifically, the
bright subsample showed systematically higher values of {\omgas}
than the faint subsample. To explain the independence of 
{\dNdX} on quasar magnitude, the difference must lie
in the incidence of systems with large {\nh}, which is 
observed to be larger in the bright subsample.

Prochaska, Herbert-Fort, \& Wolfe (2005)
argue that gravitational magnification of the background
quasars by massive halos or disks associated the foreground {\DLAs} could
account for this systematic effect, which was first detected
at the 2$\sigma$ level by \citet{murphyl04}. Whereas
obscuration by dust would cause {\omgas} to be {\em lower} in the
bright subsample, magnification by lensing due to exponential
disks is greatest for {\DLAs} with large values of {\nh} 
(\citealt*{bartelmannl96}; \citealt{mallerfp97}). 

Consequently,
the values of {\omgas} in Figure (5) may be 10-20 $\%$ too high, but
the evolution of {\omgas} with $z$ is still likely to be correct.
However, the effect could have a more important effect on
{\DLA} sample with  $z \ <$2 (\citealt{raot00}).

\section{CONCLUSIONS}

We end this review with the following question: What have we learned
from {\DLAs} that we did not know before? 
We attempt to answer this question by listing results 
judged to be robust. These are also summarized in Tables~\ref{tab:indv} 
and \ref{tab:global}, which 
describe  cosmological and local properties, respectively. Table 3
lists both the medians and the means 
in order to
show the effects of the upper limits placed on various parameters.
Specifically, the means only include positively detected quantities
while the medians include upper limits.

(1) Most of the neutral gas in the Universe in the redshift
interval $z$=[0,5] is in {\DLAs}. The cosmology and mean intensity
of extragalactic radiation are sufficiently well known to
justify the assumption of gas neutrality for {\nh}
$\ge$ {\NL}.
The close agreement between the mass per unit comoving volume
of neutral gas in {\DLAs} and visible matter in current galaxies
indicates that 
{\DLAs} comprise a significant neutral-gas
reservoir for star formation at high redshift. 

(2) The comoving density of the neutral gas, {\omgas}, declines
by a factor of two between $z$ $\approx$ 3.5 and $z$ $\approx$ 2.3.
While the evolution at 0 $<$ $z$ $<$ 2.3 is more uncertain,
{\omgas} at $z$ $\approx$ 3.5 is a factor of three higher than
at $z$ = 0.

(3) {\DLAS} are metal-poor 
at all redshifts (see Table~\ref{tab:global}),
but exhibit a metallicity ``floor'', 
[M/H] $\geq -$2.6 
(Table~\ref{tab:indv}),
indicating a different enrichment history than that of the {\lya}
forest.

%

(4) The cosmic metallicity
doubles every Gyr at $z$ $>$ 2, but the median
[M/H] is sub-solar at $z$ $\le$ 1.6.

(5) From the large [Zn/Cr] ratios
and the increase of the [Zn/Fe] and [Si/Fe]
ratios with increasing metallicity
we know that {\DLAs} exhibit evidence for depletion by dust
and that the dust content is far lower than in the Galaxy.

(6) The presence of a plateau in the [N/$\alpha$]
versus [$\alpha$/H] plane 
near [N/$\alpha$] $\approx$ $-0.7$ indicates a minimum age of
0.25 Gyr for {\DLAs}, which suggests they are
not transient objects but instead probably have ages comparable
to the Hubble time at the absorption epoch.

(7) Ionized gas in {\DLAs} exhibits a different velocity
structure than the neutral  gas, unlike the agreement between
the velocity structures of these two phases in the Galaxy. 

(8) H$_{2}$ and other
molecules are rarely  present in {\DLAs}. Studies of those systems
exhibiting H$_{2}$ absorption 
indicate the presence of an FUV radiation field with
{\jnu} $\approx$ 10$^{-19}$ {\junit}, which resembles
(a) the interstellar radiation
field in the 
Galaxy and (b) {\jnu} predicted by the {\ciis} technique.

(9) The frequency distribution of the absorption
velocity intervals, $\Delta v$, has a median of
90 {\kms}. This property cannot be reproduced
by single-disk CDM scenarios proposed so far, and is
difficult to reproduce for
sightlines passing through dwarf galaxies.
{\DLAS} with 
large values of {$\Delta v$}
exhibit a systematic absence of low values of [M/H]
and high values of {\nh}.

(10) We cannot rule out the hypothesis that
galaxies identified with {\DLAs} at $z$ $<$ 1.6 are
drawn from a cross-section weighted sample of normal galaxies;
i.e., an inflated populations of  dwarfs 
is not required.

(11) {\ciis} $\lambda$ 1335.7 absorption is detected  
in about half of randomly-selected samples of {\DLAs}. The
inferred [C II] 158 {\micron} cooling rates indicate heating
rates far in excess of those supplied by FUV background
radiation, requiring a local heat source. 
The evidence accumulated so far suggests that the likely
site of {\ciis} absorption is CNM gas.

{\noindent} Next, we 
describe
critical unsolved problems in
{\dla} research.  

{(I) {\em What is the median mass, $M_{med}$, of the dark-matter halos
containing
{\DLAs}?} 
This is the critical diagnostic for discriminating
among most hierarchical models, in which 
$M_{med}$ $<$ 10$^{9}$ {\msolar}, from 
hierarchical models with feedback or passive evolution 
models, in which $M_{med}$ $>$ 10$^{11}$ {\msolar}. 


{(II) {\em Does the {\dla} luminosity function overlap that of Lyman Break
galaxies?}
Partial Overlap is suggested by the luminosities of
the few objects detected in emission. 

{(III) {\em What are the properties of the  
interstellar gas in {\DLAs}?}
These are crucial for
understanding whether or not the gas
in which {\ciis} absorption is detected
can support a  CNM. 

{(IV) {\em Are stars forming 
in {\DLAs} when they are detected?} 



{(V) {\em How are the star formation 
and
accretion histories of {\DLAs} related?} 
The {\ciis}
technique indicates that star formation depletes the neutral gas
reservoir of {\DLAs} more rapidly than indicated by the decrease
of {\omgas} with time at $z$ $>$ 2.3. Does this require accretion
of neutral gas onto {\DLAs} at rates comparable to the star formation
rates?


{(VI) {\em What is the solution to the ``missing metals''
problem?} 
Evidence for metal-enriched
gas ejected from {\DLAs}, or for light emitted from
compact
``bulge'' regions  would help
in deciding between these hypotheses.

{(VII){\em What is the intrinsic nucleosynthetic {\rm [Si/Fe]}
ratio in {\DLAs}?} 
Are the intrinsic
abundances of {\DLAs} $\alpha$-enhanced? 

{(VIII){\em What is the cosmic metallicity of low-$z$ {\DLAs}?}
Is the column-density weighted mean metallicity of low-$z$ {\DLAs}  biased
by undersampling and by obscuration?


{(IX){\em How can we improve numerical simulations of {\DLA} evolution?}
The next steps involve more accurate modeling of star formation and
mechanical feedback. 


A major goal of {\DLA} research is 
to give a clear 
and decisive answer to the question, 
``What is a {\DLA}?'' 
Obviously 
this has not yet been accomplished.  
Rather, what we have found is that 
a significant fraction of {\DLAs} are a population
of H I layers exhibiting many of the complexities
of the ISM of the Galaxy. They clearly 
play an important role in the formation of galaxies
and undoubtedly interact with other structures in 
the high redshift Universe through a variety of 
feedback mechanisms. Observations of {\DLAs}
provide an amazingly rich data set that gives information
about galaxy formation unavailable by other means.
Specifically,  observations of {\DLAs} are the only way to study in
detail
the neutral gas that gave rise to galaxies at high
redshifts. We hope that the interplay between 
new 
observations and 
improved 
theoretical modeling will lead
to 
significant 
insights into the process of galaxy formation.

{\bf ACKNOWLEDGMENTS}

This review was written while one of us (AMW) was on
sabbatical leave at the Institute of Astronomy, Cambridge, and
AMW wishes to thank the Institute of Astronomy for 
the hospitality extended to him during his visit and
for the award of a Sackler fellowship. AMW is  particularly
grateful to Max Pettini for many valuable discussions
about our favorite mutual topic. 
AMW and JXP also wish
to thank
the Kavli Institute of Theoretical Physics, Santa Barbara,
for the hospitality extended to them during their
attendance at the Galaxy Intergalactic Medium Interactions  
program.  This material is
based on work supported by the National Science
Foundations under Grant No. 
AST 03-07824 awarded to AMW and JXP and
Grant No. 
AST-0201667 awarded to EG.

\clearpage
\begin{table} \tiny
\begin{center}
\begin{tabular}{lccccccccc}
DLA &$z_{DLA}$$^{a}$ & $z_{Ly{\alpha}}$$^{b}$       &${\theta_{b}}$$^{c}$&$b$$^{d}$&$F(Line)$$^{e}$&SFR           &SFR&Ref$^{g}$  \\
    &          &                 & arcsec       &kpc&10$^{-17}$(cgs)&Diagnostic$^{f}$ &{\smpy}& \\
\hline
\hline
0458$-$02&2.0395&2.0396&0.3$\pm$0.3&2.5$\pm$2.5&5.4$^{0.2}_{0.8}$&{\lya} &  $>$1.5&1    \\
0953$+$47A&3.407 &3.415     &$<$0.5     &$<$3.7 &0.7$\pm$0.2    &{\lya},C& 0.8$\rightarrow$7.0&2 \\
2206$-$19A&1.9205&1.9229         &1          &8.4 &26$\pm$3.0       &C & 26$\rightarrow$50&3  \\
\hline
1210$+$17&1.8918&......         &0.25          &2.1&$<$2.5        &H{$\alpha$} & $<$5.0&4  \\
1244$-$34&1.8590&......    &0.16-0.24     &1.4-2.0 &$<$0.8   &H{$\alpha$} & $<$1.6&5  \\
8 DLAs&2.095-2.615&......    &$\approx$1.5     &$\approx$10.0&$<$9.0 &H{$\alpha$} & $<$30&6  \\

\end{tabular}
\end{center}
\caption{Measurements of Emission from  Damped {\lya} Systems} \label{tab:emitters}
$^{a}${Absorption redshift of DLA}

$^{b}${{\lya} emission redshift}

$^{c}${Displacement angle of emitter (or candidate) from QSO}

$^{d}${Displacement distance of emitter (or candidate) from QSO}

$^{e}${Line flux in units of ergs cm$^{-2}$ s$^{-1}$. First 3 entries are
{\lya} and second 3 are H$\alpha$}

$^{f}${Diagnostic for obtaining SFR:{\lya} is {\lya} emission, C is FUV continuum, and
H{$\alpha$} is from H{$\alpha$} emission.}

$^{g}$Reference:{(1)~\citet{mollerff04};(2)A. Bunker (2004), priv. comm; (3)~\citet{molleretal02};
(4)~\citet{kulkarnietal01};(5)~\citet{kulkarnietal00};(6)~\citet{bunkeretal99}} 

\end{table}

\begin{table} 
\begin{center}
\begin{tabular}{lcc}
&\multicolumn{2}{c}{Redshift Interval}\\
\cline{2-3}
\cline{2-3}
Property &0.0$-$1.6&1.6$-$4.5  \\
\hline
\hline
{\dndx} &....&0.077$\pm$0.016$^{b}$ \\
10$^{3}${\omgas} &0.96$\pm$0.23$^{a}$&0.92$\pm$0.21$^{b}$  \\
log$_{10}${\rhodotz} &$< \ -$1.40&$-$0.70$\pm$0.28$^{c}$  \\
$<Z>$ &$-$0.81$\pm$0.032&$-$1.33$\pm$0.09 \\
\end{tabular}
\end{center}
\caption{Global Properties} \label{tab:global}
$^{a}$ Due to large uncertainties in individual measurements, error
in mean given by propagation of experimental errors.

$^{b}$ Because of systematic decrease in {\dNdX} and {\omgas}
with time, error in mean determined by standard deviation.

$^{c}$Computed by averaging over the ``WD low''
and other dust models discussed in \citet{wolfegp03} (see
their Table 1), and by 
assuming that both DLAs with detected and undetected
{\ciis} absorption have 
same mean SFR per unit area. If  the SFR per unit area of the non-detections
were significantly
lower, {\rhodot} would decrease by about 0.3 dex. 
\end{table}

\begin{table} \small
\begin{center}
\begin{tabular}{rccccccc}
\hline
Property & $z$ & N$^{a}$ &$\bar x$$^{b}$ & $x_{med}$$^{c}$ & $\sigma$$^{d}$ & Min$^{e}$ & Max$^{f}$\\
\hline
$\log_{10} N$(H\,I) & $z>1.6$ &199 & 20.83 & 20.60 & 0.33 & 20.30 & 21.70\\
$\lbrack$M/H] & $0.3<z<4.9$ &130 &$-1.11$&$-1.48$&$0.55$&$-2.65$&$ 0.04$\\
$\lbrack$Zn/Fe] & $0.7<z<3.3$ & 38 &  0.54 &  0.42 & 0.25 &$-0.01$&  1.05\\
$\lbrack$$\alpha$/Fe] & $0.8<z<4.7$ & 70 &  0.42 &  0.38 & 0.18 &  0.03 &  1.00\\
\delv$_{low}$$^{g}$ & $1.7<z$ & 95 & 114. &  90. &   83.7 &  16.& 430.\\
\delv$_{high}$$^{h}$ & $1.7<z$ & 75 & 209. & 190. &  113.4 &  20.& 528.\\
log$_{10}$$f$({H$_2$})& $2.0<z<3.4$&33& $-$2.22 &$< \ -$5.93 &0.82 &$<\ -$6.98 &$-$0.64  \\
log$_{10}$$\ell_c$$^{i}$ & $1.7<z<4.2$ & 57 &$-26.57$& $<-26.93 $&  0.49 & $<-27.69$&$-25.35$\\
$G_{0}$$^{j}$ & $1.7<z<4.5$ & 39 &9.6&5.4& 6.5 & $<$0.24& 23\\
log$_{10}${\ps}$^{k}$ & $1.7<z<4.2$ & 40 &$-$1.95&$-$2.20& 0.30 & $< -$ 3.55& $-$1.55\\
\hline
\end{tabular}
\end{center}
\caption{DLA Individual Summary}\label{tab:indv}
$^{a}$Number of DLAs in sample

$^{b}$Mean value. Upper limits {\em excluded} in computation.  

$^{c}$Median value. Upper limits {\em included} in computation.  

$^{d}$Sample dispersion   

$^{e}$Minimum value.   

$^{f}$Maximum value.   

$^{g}$Low-ion absorption velocity interval (in {\kms}).   

$^{h}$High-ion (C IV) absorption velocity interval (in {\kms}).   

$^{i}$ {$\ell_c$} is in units of ergs s$^{-1}$ H$^{-1}$  

$^{j}${\jnu} in units of 10$^{-19}${\junit}. Computed for WD low model
in \citet{wolfeetal04}   

$^{k}$SFR per unit area for uniform disk model (in {\smpykpc}).   
Computed for WD low model in \citet{wolfeetal04}
\end{table}

\clearpage





%

%


\bibliographystyle{araa}
\bibliography{dlarefs}


\begin{figure}
\psfig{file=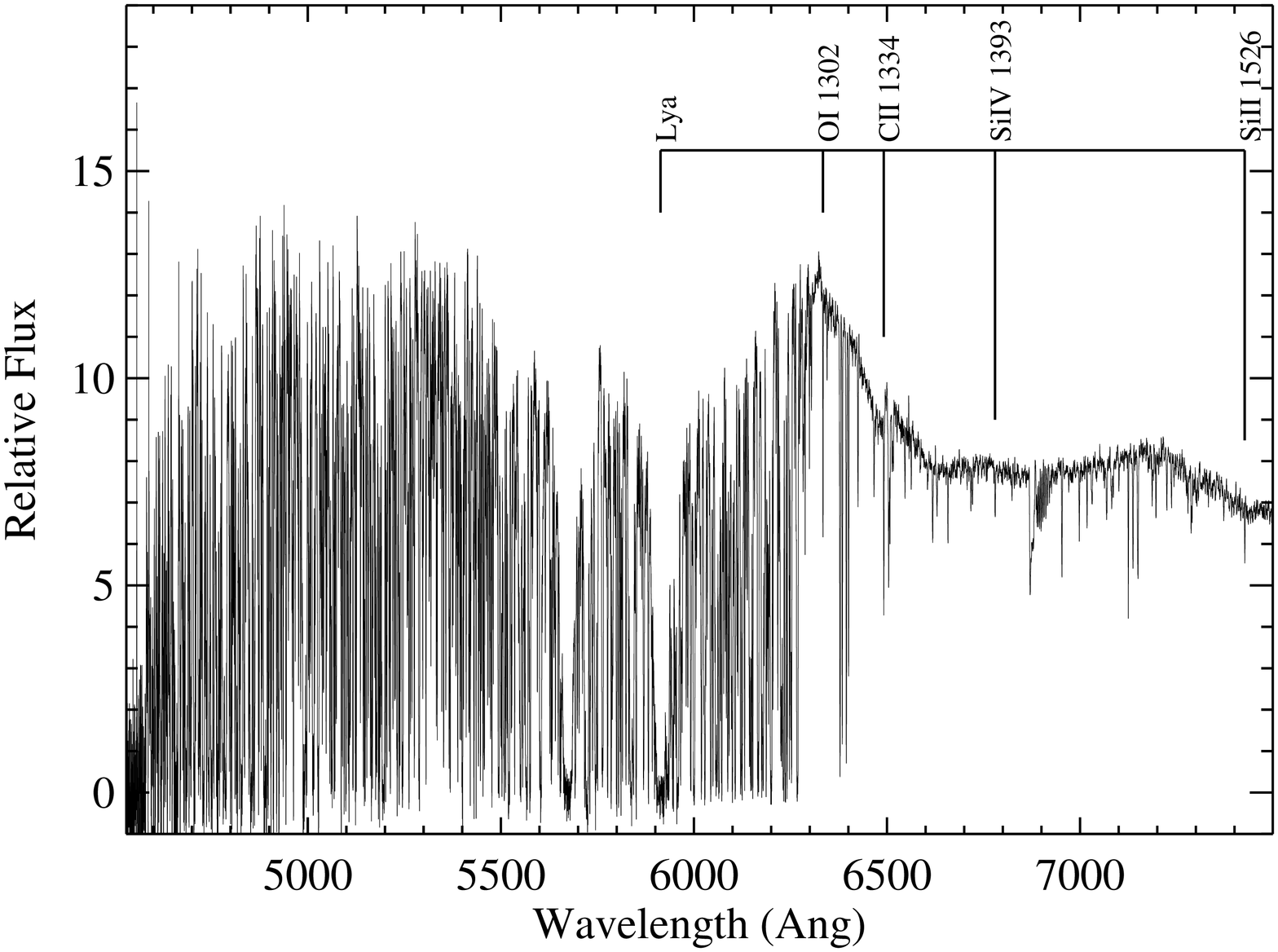,width=5in,angle=90}
\caption{Keck/ESI spectrum of QSO PSS0209+0517 showing the 
\lya\ forest, a pair of \DLAs, and a series of metal-lines.
The schematic labeling in the figure identifies several key features
for the \DLA\ at $z=3.864$. The absorption trough at $\lambda$
=5674 {\AA} corresponds to the damped {\lya} line
at $z=3.667$.}
\label{fig_lyaqso} 
\end{figure}

\begin{figure}
\psfig{file=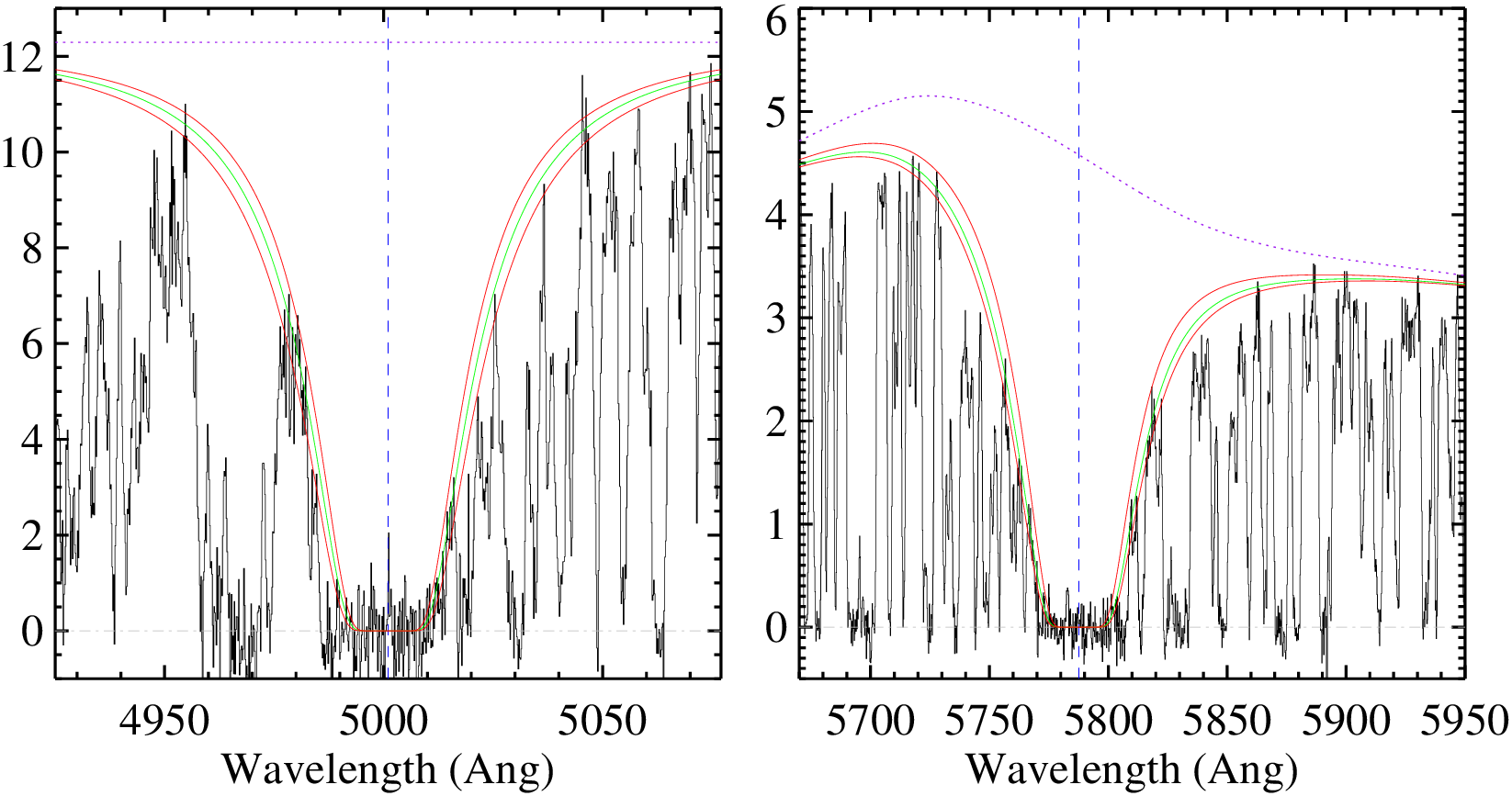,width=4in,angle=90}
\caption{Example Voigt profile fits to two \DLAs\ of the
sample from \cite{prochaskaetal03a}. The vertical dashed line indicates
the line centroid determined from metal-line transitions
identified outside the \lya\ forest.  The dotted line traces
the continuum of the QSO and the green and red lines trace
the Voigt profile solution and the fits corresponding to $1\sigma$  
changes to $\N{HI}$. The fluctuations at the bottom of the
damped {\lya} absorption troughs indicate the level of sky noise.}
\label{fig_lyafits} 
\end{figure}

\begin{figure}
\psfig{file=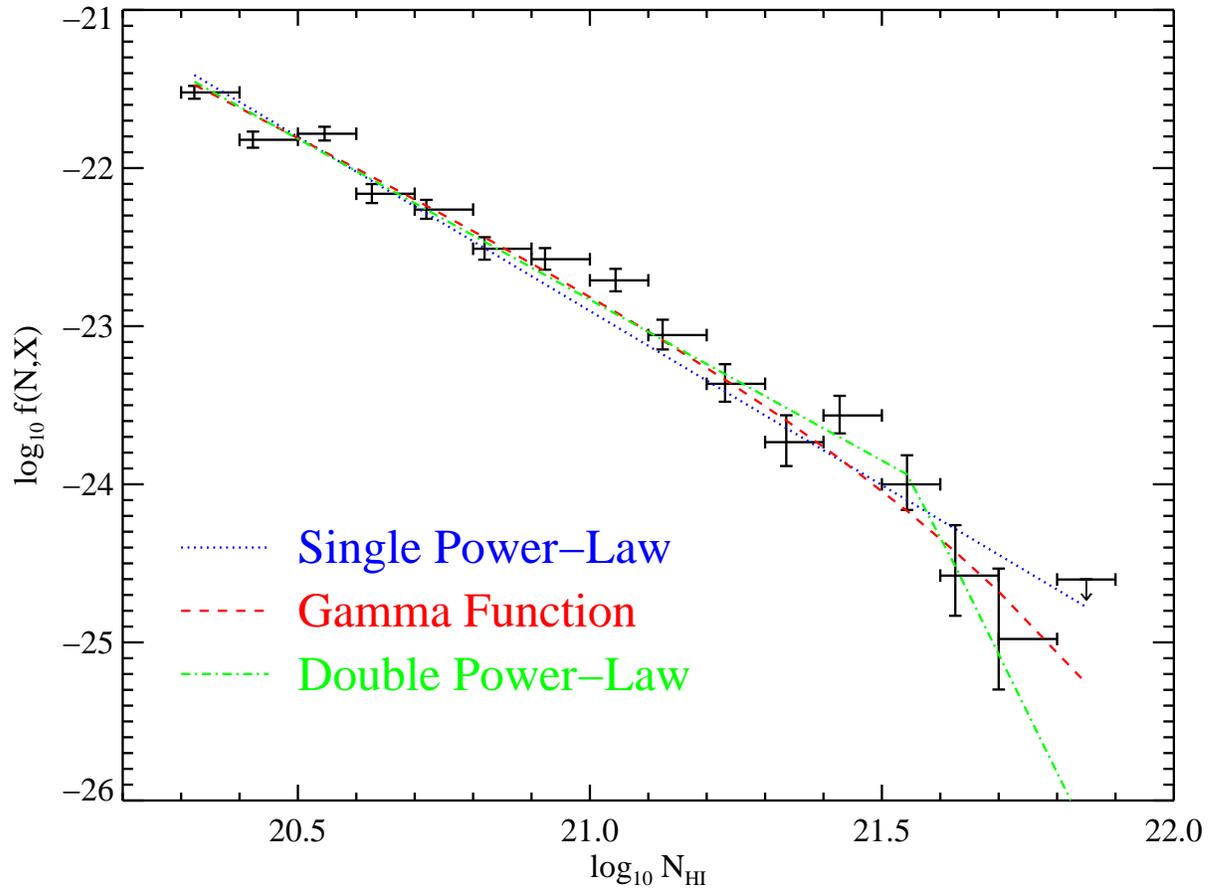,width=5in,angle=90}
\caption{The {\nh}  frequency distribution {\fNX}
determined by \cite{prochaskahw05} for
all {\DLAs} in the SDSS DR3-DR4 sample. Overplotted on the data poinst are
a single power-law, $\Gamma$ function, and
a double power-law. On the latter two are acceptable fits to the data. Plot taken
from 
\cite{prochaskahw05}.} 
\label{fig_fNvsz}
\end{figure}

\begin{figure}
\psfig{file=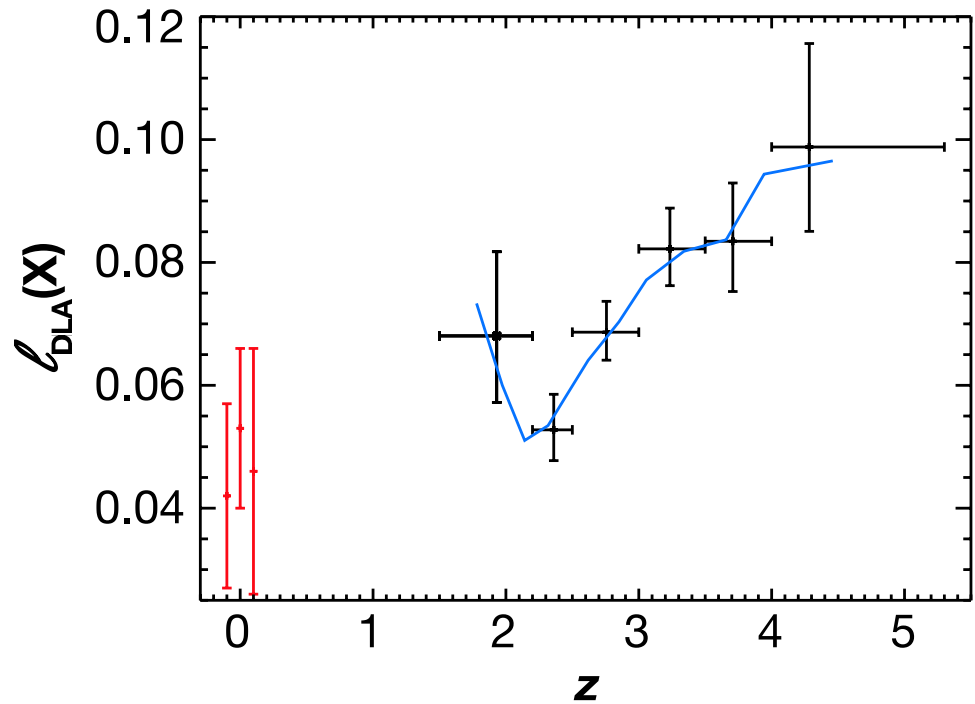,width=5in,angle=0}
\caption{Incidence of \DLAs\ per unit cosmological
distance {\dNdX} (denote by {$\ell_{DLA}$} in the figure) as a 
function of redshift.  The three data points
at $z = 0$ are all local measurements from 21\,cm observations
(\citealt*{ryanweberws03}; \citealt{zwaanetal01,rosenbergs02}).
The curve overplotted on the data traces the evaluation of {\dNdX} in a series of 0.5
Gyr intervals. Plot taken from
\cite{prochaskahw05}.}

\label{fig_nx} 
\end{figure}

\begin{figure}
\psfig{file=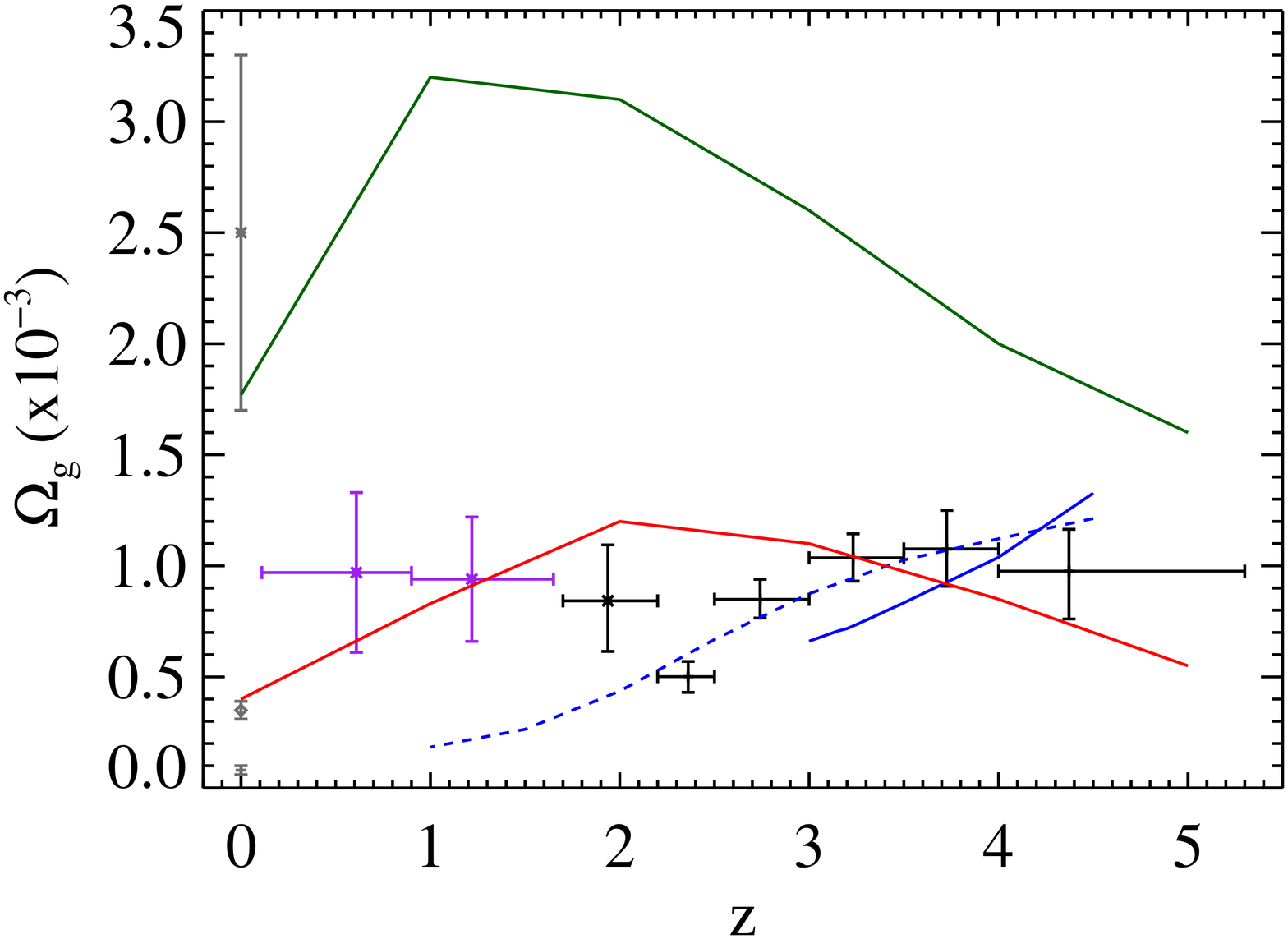,width=5in,angle=90}
\caption{Neutral gas mass density versus $z$  from
\cite{prochaskahw05}.  H I data at (a) $z$ $>$ 2.2 from SDSS DR3-4
survey, (b) 0 $<$ $z$ $<$ 1.6 from the Mg II survey of  
Rao, Turnshek, \& Nestor (2004, priv. comm.), and (c) at
$z$ = 0  (diamond) from   
\cite{fukugitahp98}. Stellar mass  density at $z$ = 0 (star) from
\cite{coleetal01} and stellar mass density of Irr galaxies ($+$ sign) from
\cite{fukugitahp98}. Theoretical curves from 
\cite{cenetal03} ({\em green}), 
\cite{somervillepf01} ({\em red}), and  
\cite{nagaminesh04a} ({\em blue:dotted} is D5 model and {\em solid} is Q5 model)} 
\label{fig_omega} 
\end{figure}

\begin{figure}
\begin{center}
\resizebox{6.3cm}{!}{\includegraphics{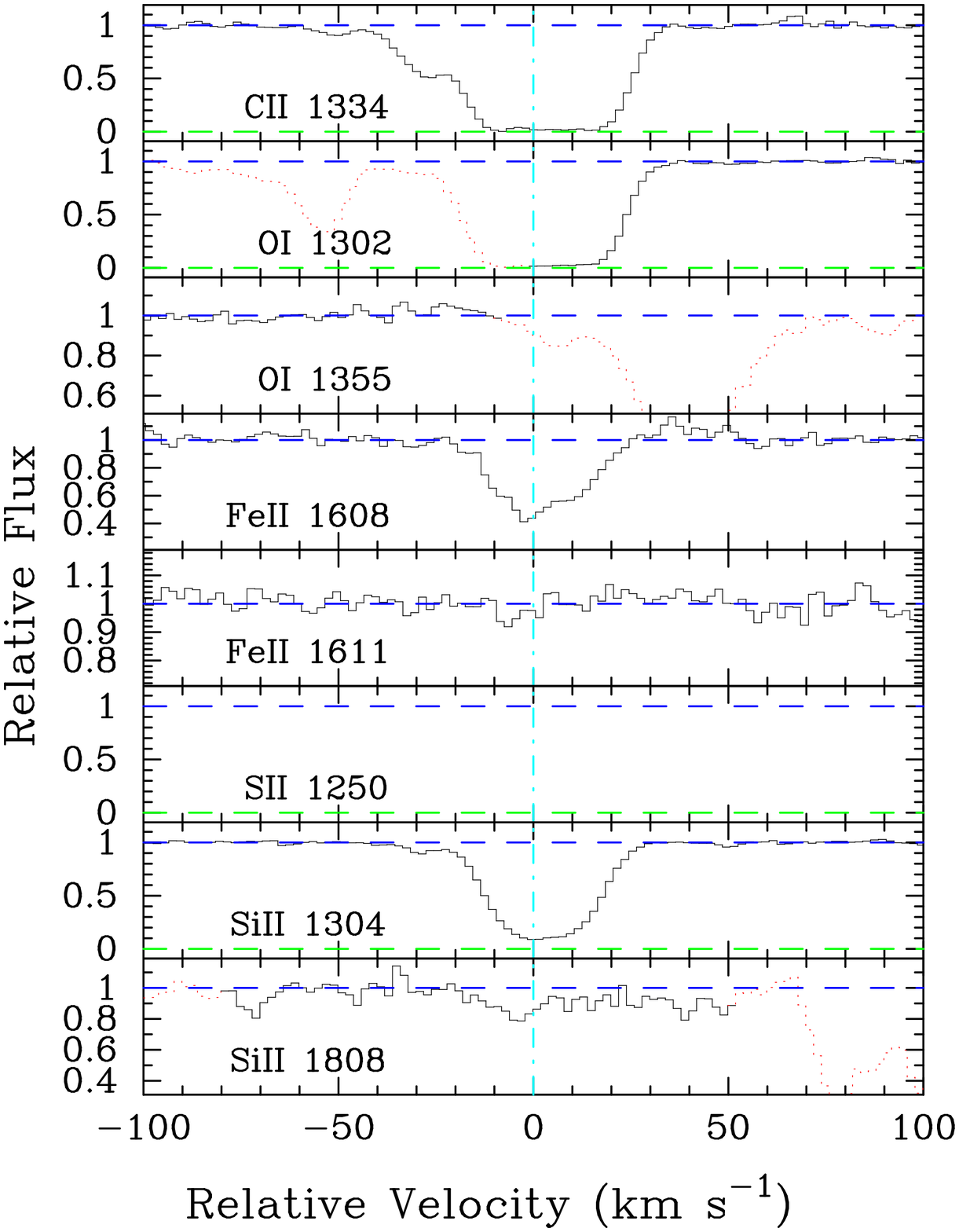}}
\resizebox{6.3cm}{!}{\includegraphics{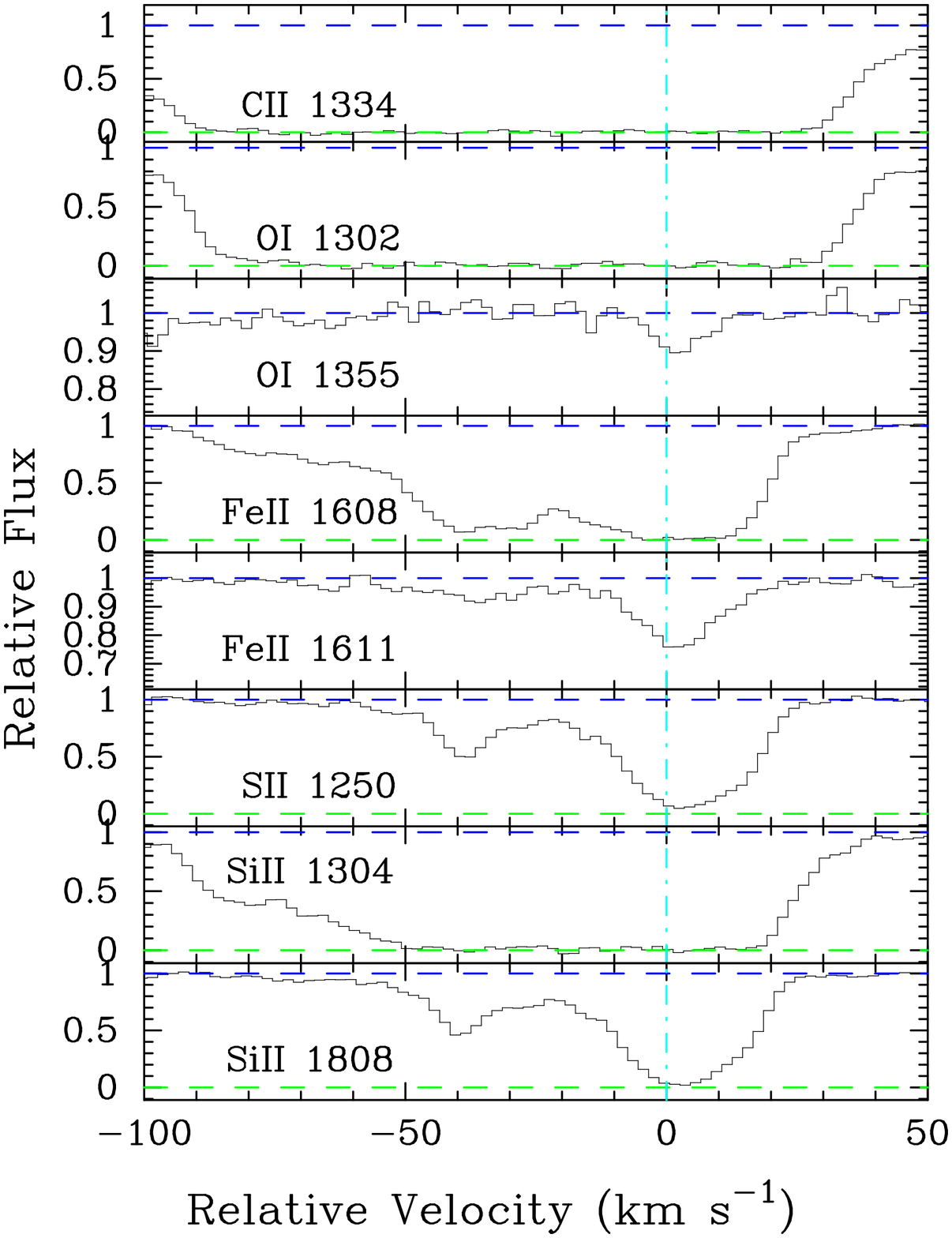}}
\caption{Velocity profiles of metal-line transitions for (a) the 
metal-poor \DLA\ at $z=3.608$ toward Q1108--07 and
(b) the metal-strong \DLA\ at $z=2.626$ toward Q0812+32.
{\em Grey} indicates the range of flux between 0 and 1. {\em Red}
lines are blends due to other transitions.
}
\label{fig_q1108}
\end{center}
\end{figure}


\begin{figure}
\psfig{file=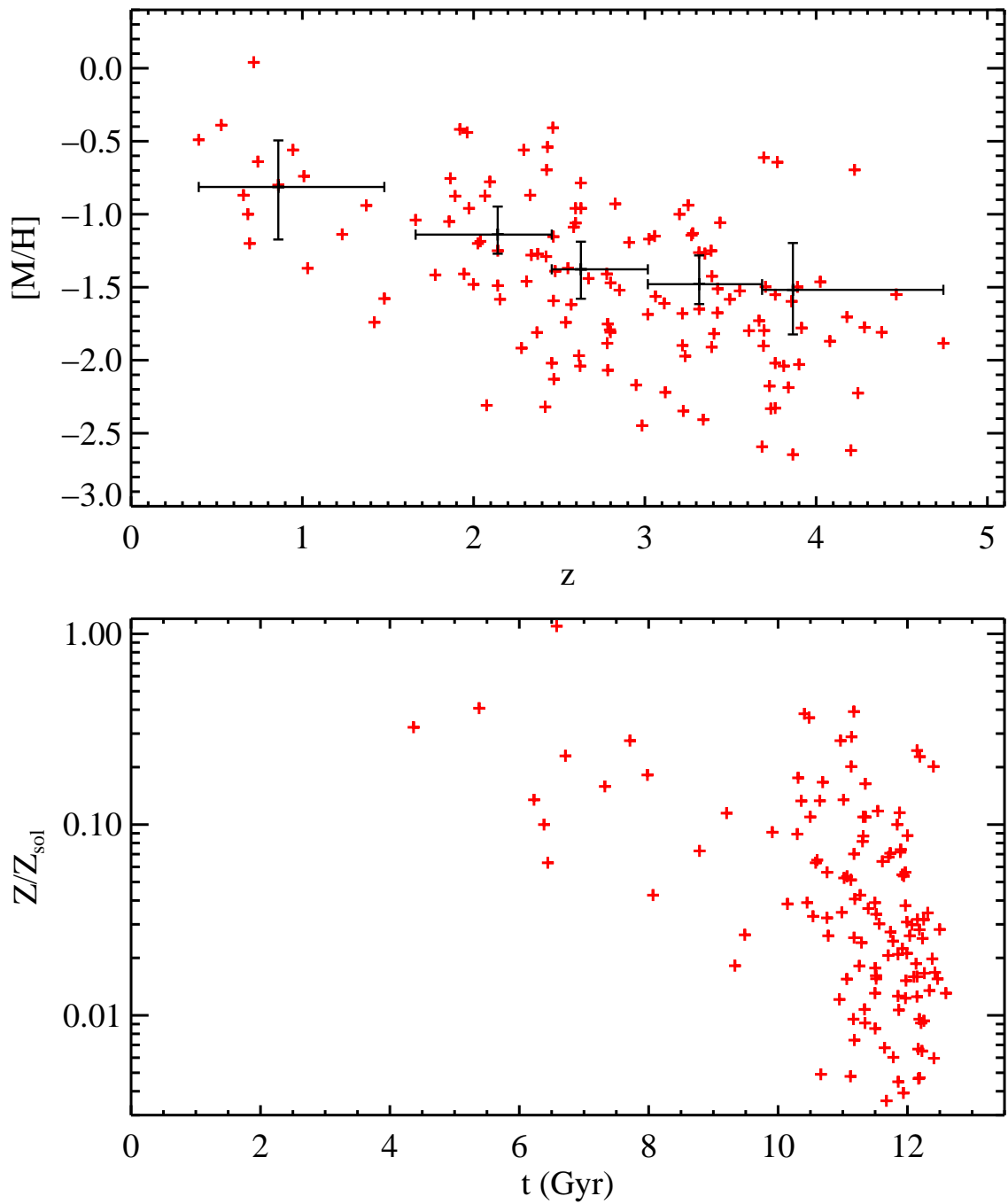,width=6in}
\caption{Current summary of the metallicity measurements of the \DLAs\
as summarized in \citet{prochaskaetal03b}, \citet{kulkarnietal05}, 
and \citet{raoetal05}.
The upper panel plots metallicities against redshift and the binned  
points indicate the cosmological mean metallicity with $95\%$\,c.l.
uncertainty.  The lower panel plots the metallicity versus 
look-back time.  
The overwhelming majority of observations are from 
$t > 10$\,Gyr.}
\label{fig_cheme} 
\end{figure}

\begin{figure}
\begin{center}
\resizebox{9.3cm}{!}{\rotatebox{-90}{\includegraphics{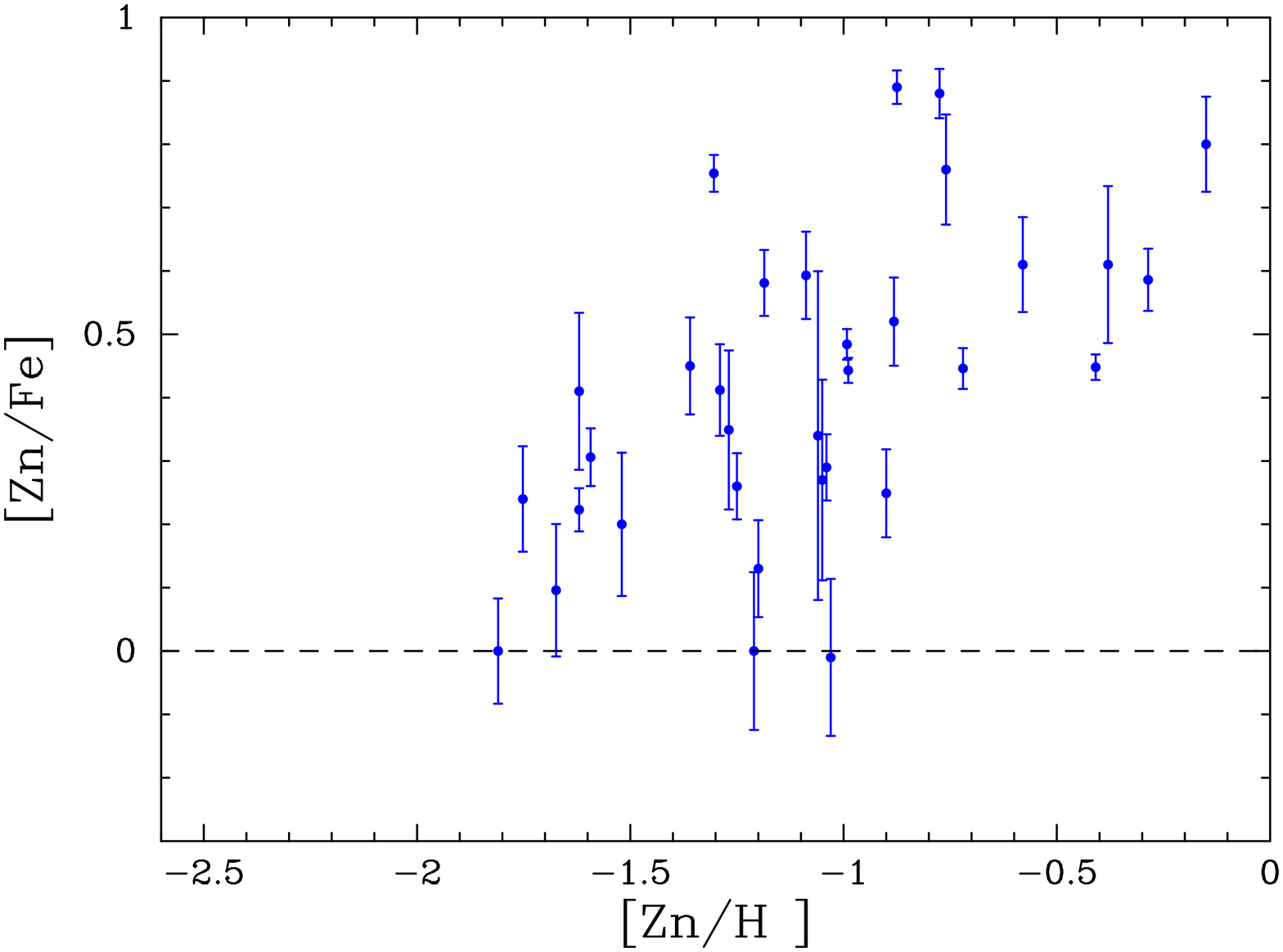}}} \\
\vspace{0.3cm} 
\resizebox{9.3cm}{!}{\rotatebox{-90}{\includegraphics{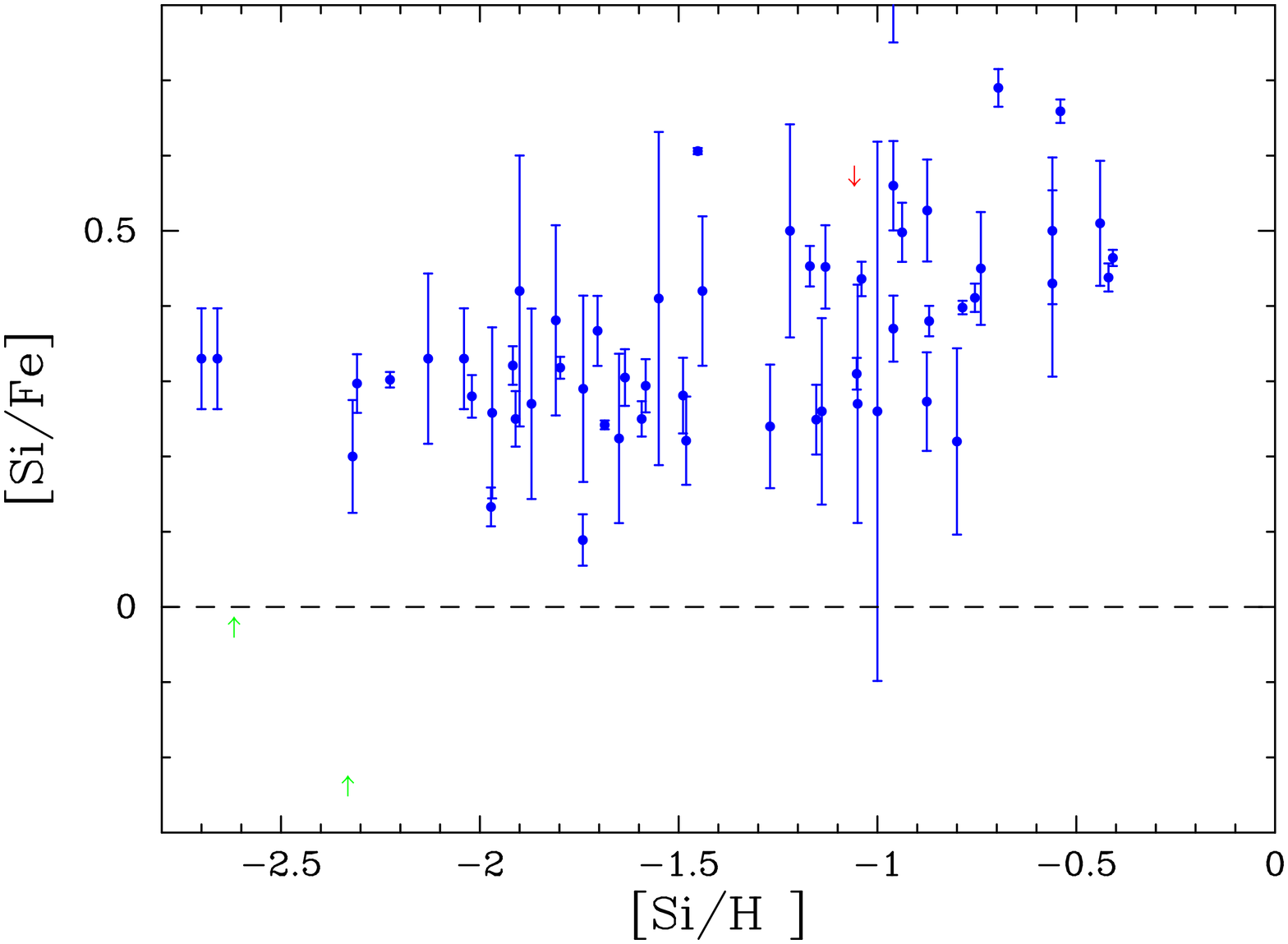}}}
\caption{Plot of (a) gas-phase [Zn/Fe] observations against Zn
metallicity for all \DLAs\ with high signal-to-noise echelle 
observations and (b)  
gas-phase [Si/Fe] observations against Si
metallicity for all \DLAs\ with high signal-to-noise echelle 
observations. \DLAS\ with upper limits to Zn were suppressed
from panel (a).}
\label{fig_ZnovFe} 
\end{center}
\end{figure}


\begin{figure}
\psfig{file=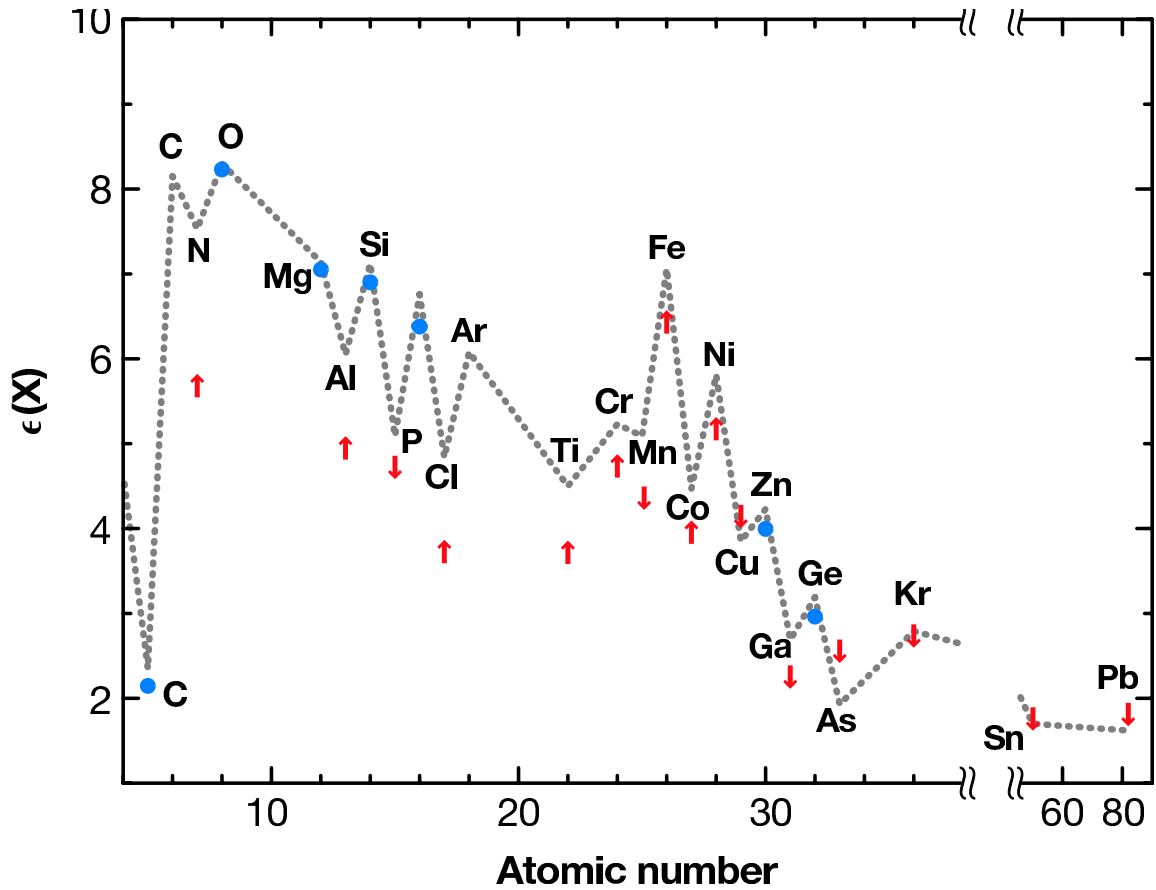,width=6in,angle=0}
\caption{Abundance pattern for the metal-strong
\DLA\ at $z=2.626$ toward Q0812+32.  
Because of the high metal abundance, [O/H] = $-$0.44,
a dust correction is necessary, and in this case a conservative
`warm halo' correction 
\citep{savages96} 
 was applied.
The dotted line traces
the Solar abundance pattern scaled to match the oxygen abundance
of the \DLA.}
\label{fig_oddeven} 
\end{figure}



\begin{figure}
\psfig{file=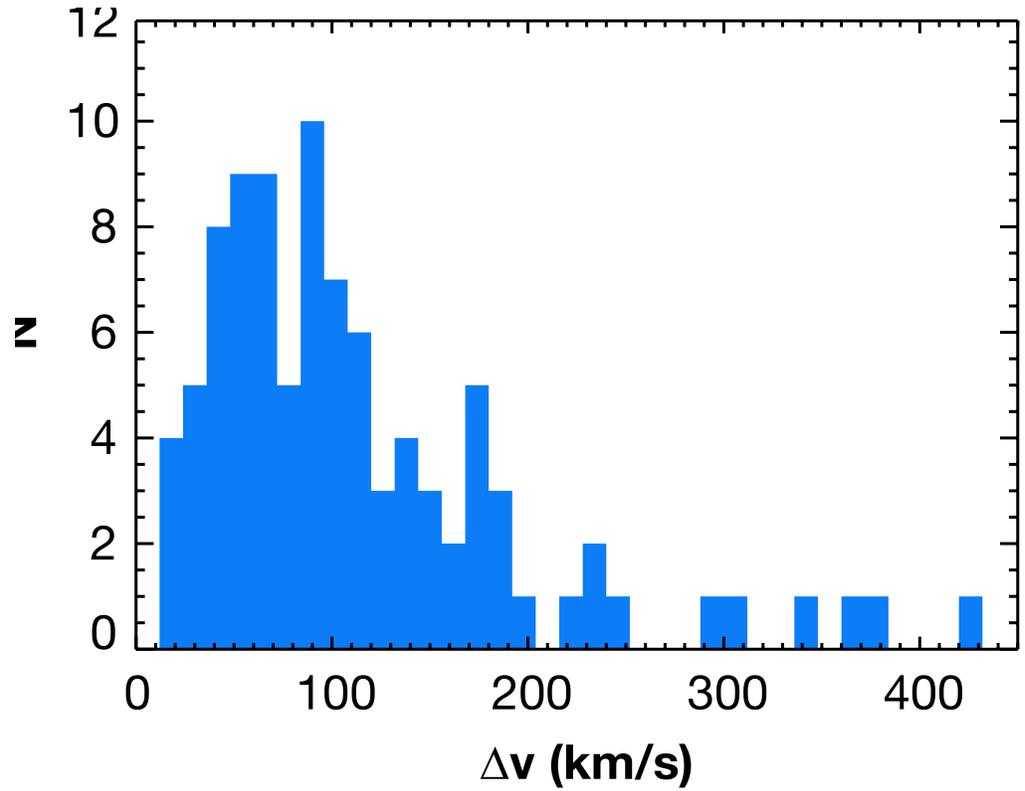,width=5in,angle=0}
\caption{Histogram of the low-ion velocity widths for the current
sample of \DLAs\ with HIRES, UVES, or ESI observations.  The median
\delv\ value is 90\kms\ and the distribution shows a significant tail
to beyond 200\kms.}
\label{fig:lowkin} 
\end{figure}

\begin{figure}
\psfig{file=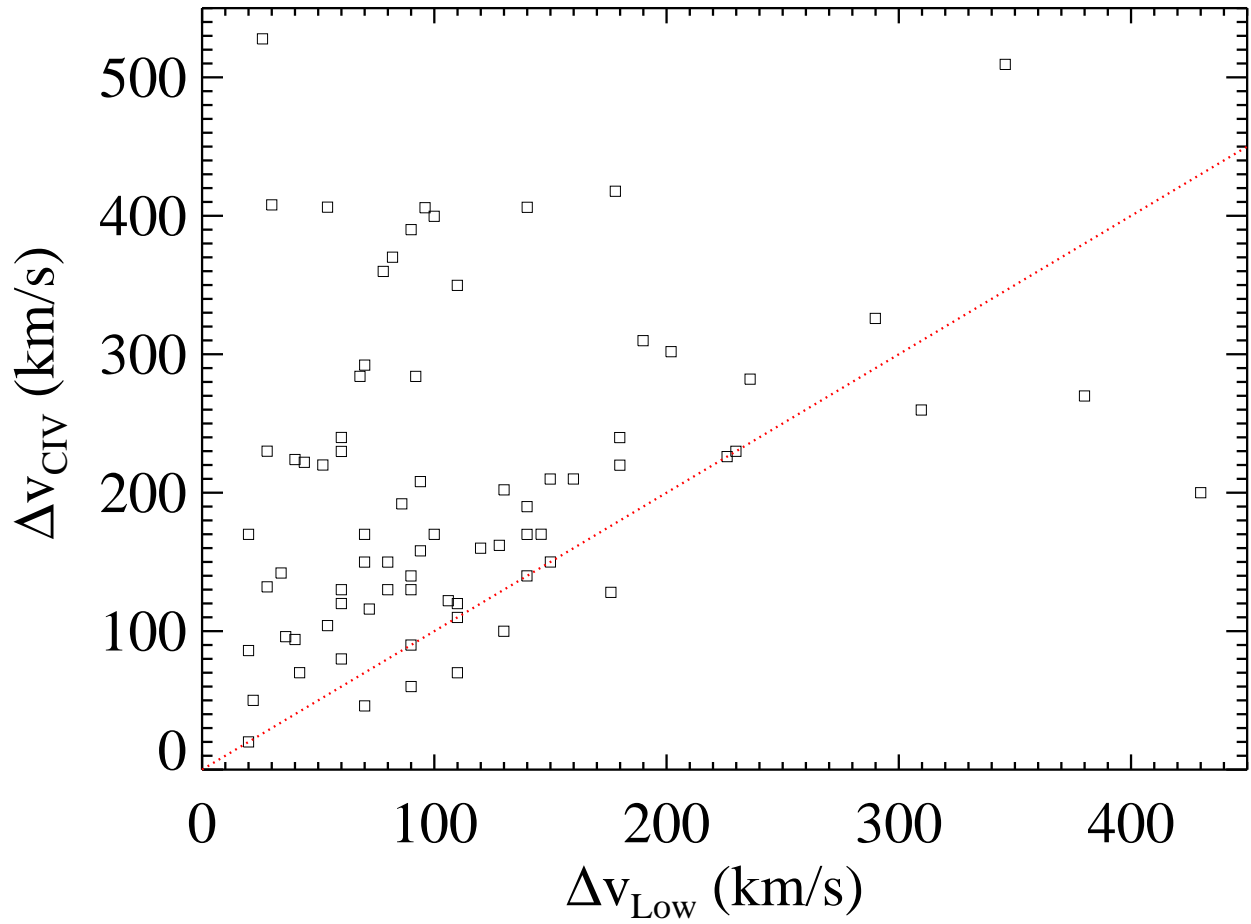,width=5in,angle=90}
\caption{Comparison of the C\,IV velocity width with the low-ion
velocity width for the current sample of \DLAs\ with HIRES or ESI
observations.  With only one or two exceptions, 
\delv$_{CIV} > \sim \Delta$v$_{Low}$.}
\label{fig:kinciv} 
\end{figure}



\begin{figure}
\psfig{file=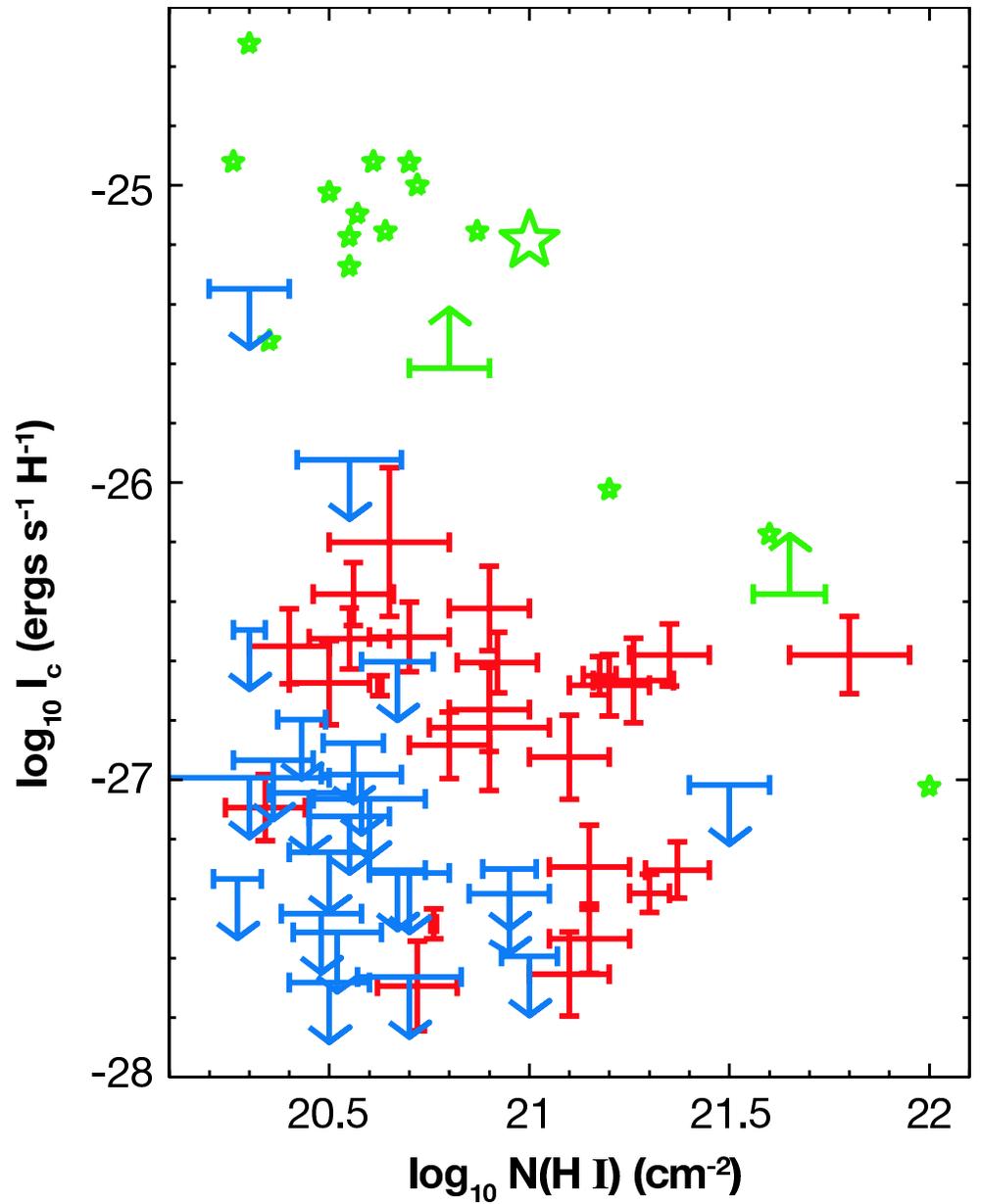,width=5in}
\caption{{\lclos} versus {\nh} for sample of 52 {\DLAs}. {\em Red}
data points are positive detections, {\em blue} are 2$\sigma$ upper limits,
and {\em green} are 2$\sigma$ lower limits. Small stars depict positive
detections 
from sightlines in the Galaxy ISM. Large star depicts 
[C II] 158 {\micron} emission rate per H atom
averaged
over the disk of the Galaxy. The latter is about 30 times
higher than the average of the DLA
detections.}
\label{fig_lcvsNHI} 
\end{figure}

\end{document}